\documentclass[aps,reprint,nofootinbib,nobibnotes,showpacs,superscriptaddress]{revtex4-1}

\usepackage[english]{babel}
	\addto\captionsenglish{%
	}

\usepackage{bm}
\usepackage{graphicx}
\usepackage{dcolumn} 
\usepackage{multirow}

\usepackage{color}
\usepackage{amssymb,amsmath,graphicx}
\usepackage{epsfig}
\usepackage[absolute,overlay]{textpos}

\usepackage[version=3]{mhchem}	
\usepackage{bm}	
\usepackage{nicefrac}	
\usepackage{mathrsfs}

\usepackage[usenames,dvipsnames]{xcolor}	
\usepackage{lineno}
\usepackage{verbatim}
\usepackage[version=3]{mhchem}
\usepackage[hidelinks]{hyperref}
\hypersetup{
    colorlinks,
    linkcolor={blue!70!black},
    citecolor={blue!70!black},
    urlcolor={blue!70!black},
    breaklinks=true
}

\usepackage{array}
\newcolumntype{L}[1]{>{\raggedright\let\newline\\\arraybackslash\hspace{0pt}}m{#1}}
\newcolumntype{C}[1]{>{\centering\let\newline\\\arraybackslash\hspace{0pt}}m{#1}}
\newcolumntype{R}[1]{>{\raggedleft\let\newline\\\arraybackslash\hspace{0pt}}m{#1}}
\newcommand{\as}[1]{\renewcommand{\arraystretch}{#1}}

\newcommand{\APC}{AstroParticule et Cosmologie, Universit\'e Paris Diderot, CNRS/IN2P3, CEA/IRFU, Observatoire de Paris, Sorbonne Paris Cit\'e, 75205 Paris Cedex 13, France}
\newcommand{\Dubna}{Joint Institute for Nuclear Research, 141980 Dubna, Russia}
\newcommand{\Genova}{Dipartimento di Fisica, Universit\`a degli Studi and INFN, 16146 Genova, Italy}

\newcommand{\Krakow}{M.~Smoluchowski Institute of Physics, Jagiellonian University, 30348 Krakow, Poland}
\newcommand{\Kiev}{Kiev Institute for Nuclear Research, 03680 Kiev, Ukraine}
\newcommand{\Kurchatov}{National Research Centre Kurchatov Institute, 123182 Moscow, Russia}
\newcommand{\Kurchatovb}{ National Research Nuclear University MEPhI (Moscow Engineering Physics Institute), 115409 Moscow, Russia}
\newcommand{\LNGS}{INFN Laboratori Nazionali del Gran Sasso, 67010 Assergi (AQ), Italy}
\newcommand{\Milano}{Dipartimento di Fisica, Universit\`a degli Studi and INFN, 20133 Milano, Italy}
\newcommand{\Perugia}{Dipartimento di Chimica, Biologia e Biotecnologie, Universit\`a degli Studi e INFN, 06123 Perugia, Italy}
\newcommand{\Peters}{St. Petersburg Nuclear Physics Institute NRC Kurchatov Institute, 188350 Gatchina, Russia}
\newcommand{\Princeton}{Physics Department, Princeton University, Princeton, NJ 08544, USA}
\newcommand{\PrincetonChemEng}{Chemical Engineering Department, Princeton University, Princeton, NJ 08544, USA}
\newcommand{\UMass}{Amherst Center for Fundamental Interactions and Physics Department, University of Massachusetts, Amherst, MA 01003, USA}
\newcommand{\Virginia}{Physics Department, Virginia Polytechnic Institute and State University, Blacksburg, VA 24061, USA}
\newcommand{\Munchen}{Physik-Department and Excellence Cluster Universe, Technische Universit\"at  M\"unchen, 85748 Garching, Germany}
\newcommand{\Lomonosov}{ Lomonosov Moscow State University Skobeltsyn Institute of Nuclear Physics, 119234 Moscow, Russia}
\newcommand{\GSSI}{ Gran Sasso Science Institute, 67100 L'Aquila, Italy}
\newcommand{\Huston}{Department of Physics, University of Houston, Houston, TX 77204, USA}
\newcommand{\Dresda}{Department of Physics, Technische Universit\"at Dresden, 01062 Dresden, Germany}
\newcommand{\Mainz}{Institute of Physics and Excellence Cluster PRISMA, Johannes Gutenberg-Universit\"at Mainz, 55099 Mainz, Germany}
\newcommand{\Honolulu}{Department of Physics and Astronomy, University of Hawaii, Honolulu, HI 96822, USA}
\newcommand{\Juelich}{Institut f\"ur Kernphysik, Forschungszentrum J\"ulich, 52425 J\"ulich, Germany}
\newcommand{\RWTH}{RWTH Aachen University, 52062 Aachen, Germany}
\newcommand{\Tubingen}{Kepler Center for Astro and Particle Physics, Universit\"{a}t T\"{u}bingen, 72076 T\"{u}bingen, Germany}
\newcommand{\Aquila}{Dipartimento di Scienze Fisiche e Chimiche, Universit\`a dell'Aquila, 67100 L'Aquila, Italy}
\newcommand{\Roma}{Present address: Dipartimento di Fisica, Sapienza Universit\`a di Roma e INFN, 00185 Roma, Italy}
\newcommand{\Napoli}{Present address: Dipartimento di Fisica, Universit\`a degli Studi Federico II e INFN, 80126 Napoli, Italy}
\newcommand{\Madrid}{Present address: Universidad Aut�noma de Madrid, Ciudad Universitaria de Cantoblanco, 28049 Madrid, Spain}
\newcommand{\Fermi}{Present address: Fermi National Accelerato Laboratory (FNAL), Batavia, IL 60510, USA}
\newcommand{\LNGSG}{Present address: INFN Laboratori Nazionali del Gran Sasso, 67010 Assergi (AQ), Italy}
\newcommand{\KFKI}{Also at: MTA-Wigner Research Centre for Physics, Department of Space Physics and Space Technology, 1121 Budapest, Hungary}

\newcommand\blfootnote[1]{%
  \begingroup
  \renewcommand\thefootnote{}\footnote{#1}%
  \addtocounter{footnote}{-1}%
  \endgroup
}
\newcommand{\spokes}{Corresponding author: spokeperson-borex@lngs.infn.it}

\usepackage{lineno}

\begin{document}

\preprint{AIP/123-QED}

\title{Simultaneous Precision Spectroscopy of {\it pp}, $^7$Be, and {\it pep} Solar Neutrinos with Borexino Phase-II}

\author{M.~Agostini}
\affiliation{\Munchen}
\author{K.~Altenm\"{u}ller}
\affiliation{\Munchen}
\author{S.~Appel}
\affiliation{\Munchen}
\author{V.~Atroshchenko}
\affiliation{\Kurchatov}
\author{Z.~Bagdasarian}
\affiliation{\Juelich}
\author{D.~Basilico}
\affiliation{\Milano}
\author{G.~Bellini}
\affiliation{\Milano}
\author{J.~Benziger}
\affiliation{\PrincetonChemEng}
\author{G.~Bonfini}
\affiliation{\LNGS}
\author{D.~Bravo$^{+}$}\blfootnote{$^{+}$ \Madrid}
\affiliation{\Milano}
\author{B.~Caccianiga}
\affiliation{\Milano}
\author{F.~Calaprice}
\affiliation{\Princeton}
\author{A.~Caminata}
\affiliation{\Genova}
\author{L.~Cappelli}
\affiliation{\LNGS}
\author{S.~Caprioli}
\affiliation{\Milano}
\author{M.~Carlini}
\affiliation{\LNGS}
\author{P.~Cavalcante$^{++}$}\blfootnote{$^{++}$ \LNGSG}
\affiliation{\Virginia}
\author{F.~Cavanna}
\affiliation{\Genova}
\author{A.~Chepurnov}
\affiliation{\Lomonosov}
\author{K.~Choi}
\affiliation{\Honolulu}
\author{L.~Collica}
\affiliation{\Milano}
\author{D.~D'Angelo}
\affiliation{\Milano}
\author{S.~Davini}
\affiliation{\Genova}
\author{A.~Derbin}
\affiliation{\Peters}
\author{X.F.~Ding}
\affiliation{\GSSI}
\affiliation{\LNGS}
\author{A.~Di Ludovico} 
\affiliation{\Princeton}
\author{L.~Di Noto}
\affiliation{\Genova}
\author{I.~Drachnev}
\affiliation{\Peters}
\author{K.~Fomenko}
\affiliation{\Dubna}
\author{A.~Formozov}
\affiliation{\Dubna}
\affiliation{\Milano}
\affiliation{\Lomonosov}
\author{D.~Franco}
\affiliation{\APC}
\author{F.~Gabriele}
\affiliation{\LNGS}
\author{C.~Galbiati}
\affiliation{\Princeton}
\affiliation{\GSSI}
\author{M.~Gschwender}
\affiliation{\Tubingen}
\author{C.~Ghiano}
\affiliation{\LNGS}
\author{M.~Giammarchi}
\affiliation{\Milano}
\author{A.~Goretti$^{++}$}
\affiliation{\Princeton}
\author{M.~Gromov}
\affiliation{\Lomonosov}
\affiliation{\Dubna}
\author{D.~Guffanti}
\affiliation{\GSSI}
\affiliation{\LNGS}
\author{T.~Houdy}
\affiliation{\APC}
\author{E.~Hungerford}
\affiliation{\Huston}
\author{Aldo~Ianni}
\affiliation{\LNGS}
\author{Andrea~Ianni}
\affiliation{\Princeton}
\author{A.~Jany}
\affiliation{\Krakow}
\author{D.~Jeschke}
\affiliation{\Munchen}
\author{S.~Kumaran}
\affiliation{\Juelich}
\affiliation{\RWTH}
\author{V.~Kobychev}
\affiliation{\Kiev}
\author{G.~Korga$^{*}$}\blfootnote{$^{*}$ \KFKI}
\affiliation{\Huston}
\author{T.~Lachenmaier}
\affiliation{\Tubingen}
\author{M.~Laubenstein}
\affiliation{\LNGS}
\author{E.~Litvinovich}
\affiliation{\Kurchatov}
\affiliation{\Kurchatovb}
\author{P.~Lombardi}
\affiliation{\Milano}
\author{L.~Ludhova}
\affiliation{\Juelich}
\affiliation{\RWTH}
\author{G.~Lukyanchenko}
\affiliation{\Kurchatov}
\author{L.~Lukyanchenko}
\affiliation{\Kurchatov}
\author{I.~Machulin}
\affiliation{\Kurchatov}
\affiliation{\Kurchatovb}
\author{G.~Manuzio}
\affiliation{\Genova}
\author{S.~Marcocci$^{\dagger\dagger}$}\blfootnote{$^{\dagger\dagger}$ \Fermi}
\affiliation{\GSSI}
\author{J.~Maricic}
\affiliation{\Honolulu}
\author{J.~Martyn}
\affiliation{\Mainz}
\author{E.~Meroni}
\affiliation{\Milano}
\author{M.~Meyer}
\affiliation{\Dresda}
\author{L.~Miramonti}
\affiliation{\Milano}
\author{M.~Misiaszek}
\affiliation{\Krakow}
\author{V.~Muratova}
\affiliation{\Peters}
\author{B.~Neumair}
\affiliation{\Munchen}
\author{M.~Nieslony}
\affiliation{\Mainz}
\author{L.~Oberauer}
\affiliation{\Munchen}
\author{V.~Orekhov}
\affiliation{\Kurchatov}
\author{F.~Ortica}
\affiliation{\Perugia}
\author{M.~Pallavicini}
\affiliation{\Genova}
\author{L.~Papp}
\affiliation{\Munchen}
\author{\"O.~Penek}
\affiliation{\Juelich}
\affiliation{\RWTH}
\author{L.~Pietrofaccia}
\affiliation{\Princeton}
\author{N.~Pilipenko}
\affiliation{\Peters}
\author{A.~Pocar}
\affiliation{\UMass}
\author{A.~Porcelli}
\affiliation{\Mainz}
\author{G.~Raikov}
\affiliation{\Kurchatov}
\author{G.~Ranucci}
\affiliation{\Milano}
\author{A.~Razeto}
\affiliation{\LNGS}
\author{A.~Re}
\affiliation{\Milano}
\author{M.~Redchuk}
\affiliation{\Juelich}
\affiliation{\RWTH}
\author{A.~Romani}
\affiliation{\Perugia}
\author{N.~Rossi$^{+\dagger}$}\blfootnote{$^{+\dagger}$ \Roma}
\affiliation{\LNGS}
\author{S.~Rottenanger}
\affiliation{\Tubingen}
\author{S.~Sch\"onert}
\affiliation{\Munchen}
\author{D.~Semenov}
\affiliation{\Peters}
\author{M.~Skorokhvatov}
\affiliation{\Kurchatov}
\affiliation{\Kurchatovb}
\author{O.~Smirnov}
\affiliation{\Dubna}
\author{A.~Sotnikov}
\affiliation{\Dubna}
\author{L.F.F.~Stokes}
\affiliation{\LNGS}
\author{Y.~Suvorov$^{\dagger}$}\blfootnote{$^{\dagger}$ \Napoli}
\affiliation{\LNGS}
\affiliation{\Kurchatov}
\author{R.~Tartaglia}
\affiliation{\LNGS}
\author{G.~Testera}
\affiliation{\Genova}
\author{J.~Thurn}
\affiliation{\Dresda}
\author{E.~Unzhakov}
\affiliation{\Peters}
\author{F.~Villante}
\affiliation{\LNGS}
\affiliation{\Aquila}
\author{A.~Vishneva}
\affiliation{\Dubna}
\author{R.B.~Vogelaar}
\affiliation{\Virginia}
\author{F.~von~Feilitzsch}
\affiliation{\Munchen}
\author{S.~Weinz}
\affiliation{\Mainz}
\author{M.~Wojcik}
\affiliation{\Krakow}
\author{M.~Wurm}
\affiliation{\Mainz}
\author{O.~Zaimidoroga}
\affiliation{\Dubna}
\author{S.~Zavatarelli}
\affiliation{\Genova}
\author{K.~Zuber}
\affiliation{\Dresda}
\author{G.~Zuzel} 
\affiliation{\Krakow}

\collaboration{Borexino Collaboration$^{**}$}\blfootnote{$^{**}$ \spokes}
\noaffiliation{}
\collaboration{  This work is dedicated to the  memory of Simone Marcocci, a young brilliant scientist, a valuable collaborator  and  our  great friend. }
\noaffiliation{}

\date{\today}

\begin{abstract}
  We present the simultaneous measurement of the interaction rates $R_{pp}$, $R_{\rm {Be}}$, $ R_{pep}$ of $pp$, $^7$Be, and $pep$ solar neutrinos performed
  with a global fit to the Borexino data in an extended energy range (0.19 -- 2.93)\,MeV  with particular attention to details of the analysis methods. This result was obtained
  by analyzing 1291.51\,days of Borexino Phase-II data, collected
  after an extensive scintillator purification campaign. Using 
counts per day (cpd)/100\,ton as unit, 
we find $R_{pp}$\,=\,134\,$\pm$\,10\,({\it stat})\,$^{+6}_{-10}$\,({\it sys)}, $R_{Be}$\,=\,48.3\,$\pm$\,1.1\,({\it stat})\,$^{+0.4}_{-0.7}$\,({\it sys}); and 
$R_{pep}^{\rm{HZ}}$\,= $2.43 \pm 0.36\,({\it stat})^{+0.15}_{-0.22}\,({\it sys})$ assuming the  interaction rate $R_{\rm{CNO}}$ of CNO-cycle solar neutrinos according
to the prediction of the high metallicity Standard Solar Model, and $R_{pep}^{\rm{LZ}}$\,= $2.65 \pm 0.36\,({\it stat})^{+0.15}_{-0.24}\,({\it sys})$ according to that of the
low metallicity model.


  An upper limit  $R_{\rm{CNO}}$\,$<$\,8.1\,cpd/\,100\,ton (95\%\,C.L.) is obtained by setting in the fit a constraint on the ratio $R_{pp}/R_{pep}$
  (47.7 $\pm$ 0.8 cpd/100\,ton
or 47.5 $\pm$ 0.8 cpd/100\,ton according to   the  high or low metallicity hypothesis).


\end{abstract}

\maketitle

\section{Introduction}
Solar neutrinos  produced in electron flavour $(\nu_e)$ in fusion reactions occurring in the Sun provide a unique and direct way to study the interior of our star.
The main contribution  to the solar luminosity ($\sim$99\%) comes from reactions belonging to the  $pp$ chain, while the CNO cycle is expected to play a sub-dominant
role~\cite{bib:TheoryA, *bib:TheoryB}. 

The solar neutrino $(\nu)$ spectrum, as predicted by the Standard Solar Model (SSM)~\cite{bib:Bahcall, bib:Carlos2017}, is dominated by the low-energy neutrinos
produced in the  primary $pp$ reaction (E$_{\nu}$\,$<$\,0.42\,MeV) and it extends up to $\sim$18.8\,MeV (maximum energy of the $hep$ $\nu$s). It also features
two mono-energetic lines from $^7$Be $\nu$s (E$_\nu$\,=\,0.384\,MeV and 0.862\,MeV) and one mono-energetic line from $pep$ $\nu$s (E$_\nu$\,=\,1.44\,MeV). 
Neutrinos from the CNO cycle are expected to have a continuous energy spectrum extending up to 1.74\,MeV. The spectrum of
$^8$B $\nu$s is also continuous and it ends up
at about 16.5\,MeV.

The 50-year-long experimental effort to study solar neutrinos~\cite{bib:GallexA,*bib:GallexB,*bib:Sage, *bib:Homestake, *bib:Kamiokande, *bib:SK, *bib:SNO, bib:BxBe7}
has been extremely rewarding both in terms of solar physics, by confirming the SSM predictions~\cite{bib:Carlos2017}, and in terms of particle physics, by giving a
substantial contribution to the discovery of neutrino flavour oscillations~\cite{bib:Nobel2015},~\cite{bib:Nobel2002}.
The present-day precision spectroscopy of solar neutrinos aims
 at studying the details of their energy spectrum by disentangling the contributions from the different reactions ($pp$ chain $\nu$s, namely $pp$, $^7$Be,
 $pep$, $^8$B, and $hep$ $\nu$s, and CNO cycle $\nu$s). 

 On the one hand, if the SSM predictions of solar fluxes  $\phi$ are assumed, measuring the solar neutrino interaction rates  $R$ for different reactions helps to pin down the
 electron-flavour neutrino survival probability $P_{ee}$ for different energies
(that is the probability that $\nu_e$s  do not undergo flavor oscillations while travelling from their
Sun production point to the detector). Consequently, it probes the predictions of the MSW-LMA model~\cite{bib:MSWA, *bib:MSWB} and
can set constraints on possible deviations, {\it e.g.} due to non--standard interactions (NSI)~\cite{bib:NSIa, *bib:NSIb, *bib:NSIc, *bib:NSId}.

On the other hand, if the neutrino oscillation parameters are assumed, the study of specific components of the solar neutrino spectrum can cross-check the SSM predictions.
In particular, the  experimental determination of the fluxes $\phi$ of $^7$Be, $^8$B or CNO neutrinos, which are the most sensitive ones to the solar metallicity
(the abundance of the elements heavier than He in the Sun), can help to settle the question of high (HZ) versus low (LZ) metallicity~\cite{bib:Carlos2017}.

The Borexino experiment has recently reported a comprehensive measurement of the solar neutrino spectrum from the whole $pp$ nuclear fusion chain in the energy range
of (0.19 -- 16)\,MeV. These results are presented in~\cite{bib:BxNaturePhase2} together with their physical implications. They include the updated values of the neutrino survival
probability $P_{ee}$ as a function of the neutrino energy, the first direct measurement of the ratio $\mathcal{R}$ between the $^3$He\,$+$\,$^4$He ($pp$-II) and
the $^3$He\,$+$\,$^3$He ($pp$-I) branches of the $pp$ chain obtained by combining our results on the $^7$Be and $pp$ $\nu$s, and finally a preference for the HZ-metallicity
choice in the SSM.

In this paper we present  the details of the analysis of the data belonging to the lowest part of the energy spectrum which extends from 0.19 to 2.93 MeV.
This Low Energy Region (LER) is used to extract the interaction rates $R_{pp}$, $R_{{\rm Be}}$, $R_{pep}$, as well as to set the limit on $R_{{\rm CNO}}$.
The analysis of the data from the so-called High Energy Region (HER) from 3.2 to 16\,MeV, where our sensitivity to $^8$B $\nu$s is maximized and from 11 to 20\,MeV
energy region, in which the first Borexino limit on $hep$ $\nu$s is set, is discussed in~\cite{bib:BxB8Phase2}.

While our previous measurements of the  $pp$~\cite{bib:Bxpp}, $^7$Be~\cite{bib:BxBe7},  $pep$ ~\cite{bib:Bxpep}, and $^8$B~\cite{bib:BxB8} $\nu$s were
obtained separately by analyzing data in restricted energy ranges, the results of \cite{bib:BxNaturePhase2} provide a unified analysis over the interval covering
the LER and HER. The experience from the previous analyses in different energy intervals, each of them having specific difficulties, was fundamental in
the process of building up the comprehensive understanding of our data and of  the detector response across the combined energy interval as a whole. In addition,
other important elements of the measurement are: an accurate calibration campaign~\cite{bib:BxCalib} in the energy interval ranging from 0.15 to 9\,MeV carried
out by  deploying  several radioactive sources inside the detector,
a detailed Monte Carlo (MC) simulation fine-tuned to reproduce the calibration
data simultaneously at low and at high energies~\cite{bib:BxMC}, and the use of data-processing and data-selection as well as background-rejection tools common
to the whole energy range.

The unified analysis approach in the LER, described in this work, together with a larger exposure and a reduction of the most relevant backgrounds
in the Phase-II lead
to a significant improvement of the accuracy of our previous  Phase-I results about the  $R_{\rm{Be}}$ (from 4.8\% to 2.7\%) and $R_{pep}$
(from 21.6\% to 17.4/16.3\%, depending on the HZ/LZ-SSM assumption, respectively). For $R_{pp}$ the improvement is smaller, from the precision of 11.4\% to 10.6\%.

\section{The Borexino detector and the data selection}
\label{sec:det}

The Borexino experiment is located at the Laboratori Nazionali del Gran Sasso in Italy.
The core of the detector~\cite{bib:BxDet} is 278\,ton of ultra-pure organic liquid scintillator, 
namely PC (pseudocumene, 1,2,4-trimethylbenzene) as a solvent and 1.5\,g/l of  fluor PPO (2,5-diphenyloxazole) as a solute,
contained in a 125\,$\mu$m-thick nylon  Inner Vessel (IV) of 4.25\,m radius, surrounded by  nominally 2212 8-inch ETL 9351 photomultipliers (PMTs).
Since the beginning of the data taking, we observed  a  slow PMT failure rate over  time. As a reference, the number of working channels was 1769 at
the beginning of the data-taking period considered
in this work while it was 1383 at its end.

Neutrinos of any flavour interact by elastic scattering with electrons, whose recoil produces scintillation light ($\sim$500\,photoelectrons/MeV/2000 PMTs). The  density
of target electrons in the scintillator is $(3.307 \pm 0.003) \times 10^{31}$/ 100\,ton. A non-scintillating buffer fills the space between the  IV and a Stainless-Steel Sphere
(SSS) of 6.85\,m radius, which supports the PMTs. 
The buffer liquid is further divided in two regions  by another nylon vessel of radius 5.5\,m which prevents radon emanating from the SSS and the PMTs
to enter the core of the detector. 
The entire detector is enclosed in a cylindrical tank filled with ultra-pure water and instrumented with 208 PMTs, acting as an active Cherenkov muon veto and as a passive
shield against external $\gamma$s and neutrons.


The present analysis is based on the data collected between December 14$^{\rm{th}}$, 2011 to May 21$^{\rm{st}}$, 2016, which corresponds to an exposure
of 1291.51\,days $\times$ 71.3\,t ($\sim$\,1.6 times the exposure used in \cite{bib:BxBe7}). This period belongs to the so-called Borexino Phase-II, which started  after an
extensive purification campaign of the scintillator with 6 cycles of closed-loop water extraction, which has significantly reduced the radioactive
contaminants: $^{238}$U $<$\,9.4\,$\times$\,10$^{-20}$\,g/g (95\% C.L.),  $^{232}$Th $<$\,5.7\,$\times$\,10$^{-19}$\,g/g (95\% C.L.), $^{85}$Kr and $^{210}$Bi
reduced respectively by a factor $\sim$4.6 and $\sim$2.3 (see this work). 

The expected solar $\nu$s interaction rate in Borexino ranges from few to $\sim$100 cpd/100\,ton depending on the neutrino component. Together with the lack
of directionality  information from the scintillation light, this low rate demands a high  detector radio-purity, a deep understanding of the backgrounds, and an accurate
modelling of the detector response.

The position and pulse-shape of each event are reconstructed by exploiting the number of detected photons and their detection times. 
The information about the event energy is carried by the number of detected photoelectrons 
  or just the number of hit PMTs, as in our energy range the PMTs mainly work in a single photoelectron regime.
In detail, we define different energy estimators:
 $N_p$ which is the total number of hit PMTs in the event or $N_p^{dt_{1(2)}}$, the number of hit PMTs happening within a fixed 
 time interval of 230 (400)\,ns; 
$N_h$ the number of detected hits, including multiple hits on the same PMT and,  finally 
$N_{pe}$, the total charge collected by each PMT anode, that is the number of photoelectrons, p.e.
 As it will be detailed in section Sec.~\ref{sec:Oleg}, the energy is not reconstructed meaning that, during the analysis procedure,
 we do not convert the values of the energy estimator into the  event energy. On the contrary, we build the prediction of the measured variables
 transforming the theoretical event energy into the corresponding value of a given energy estimator.
As a reference, at 1\,MeV, the energy and position reconstruction resolutions are $\sim$\,50\,keV and $\sim$\,10\,cm, respectively.
The trigger threshold is $N_p$\,$>$\,20  in a 100\,ns time window, which corresponds to $\sim$50\,keV.

To account for the variation in the number of working channels as a function of time, in the analysis and simulation procedures, all the energy estimators
are normalised to a fixed number $N_{tot}$ of PMTs (typically $N_{tot}$ = 2000 PMTs) \cite{bib:BxLong}
through the relation $N_{p,h,pe} = N_{p,h,pe}^m \cdot N_{tot}/N^{\prime}(t)$, with $N_{p,h,pe}^m$ being the measured value of the energy estimator and
$N^{\prime}(t)$ is the  time-dependent number of working PMTs.

Events in the entire LER are selected using the same cuts described in~\cite{bib:Bxpp}: 
we remove internal (external) muons~\cite{bib:BxMuon} and we apply a 300 (2)\,ms veto 
to suppress cosmogenic backgrounds. The total dead-time introduced by these vetoes is 1.5\%. We remove $^{214}$Bi -- $^{214}$Po fast coincidences from
the $^{238}$U chain and unphysical noise events. The fraction of good events removed by these cuts, estimated using MC simulations~\cite{bib:BxMC} and
calibration data~\cite{bib:BxCalib}, is $\sim$0.1\%.
Background from sources external to the scintillator (nylon vessel, SSS, and PMTs) is reduced with a fiducial volume (FV) cut, which selects the innermost region
of the scintillator (71.3\,ton), contained within the radius {\it R}\,$<$2.8\,m and the vertical coordinate  -1.8\,$<$\,z$<$\,2.2\,m.

\section{Background}
\label{sec:bgr}

The residual background, after the application of the described selection cuts, is mainly due to  radioactive isotopes contaminating the scintillator itself, such
as $^{14}$C ($\beta^-$ decay, Q\,=\,0.156\,MeV, $\tau$ = 8270\,years), $^{210}$Po ($\alpha$ decay, E$_{\alpha}$\,=\,5.3\,MeV, $\tau$ = 200\,days, originating a
scintillation light signal quenched by a factor $\sim$10), $^{85}$Kr ($\beta^-$ decay, Q\,=\,0.687\,MeV, $\tau = 15.4$\,years),
and $^{210}$Bi ($\beta^-$ decay, Q\,=\,1.16\,MeV, $\tau = 7.23$\,days), a relatively short lived daughter of $^{210}$Pb ($\beta^-$ decay, Q\,=\,0.063\,MeV, $\tau = 32.2$\,years). 
The lowest energy region (below 0.3\,MeV), which is most sensitive to $pp$ $\nu$s, contains an additional background
due to the {\it pile-up} of uncorrelated events (mostly $^{14}$C, external background primarly due to radioactive
contaminants of the SSS and PMTs, and $^{210}$Po~\cite{bib:Bxpp, bib:BxMC}).
The energy region sensitive to $pep$ and CNO $\nu$s (between $\sim$1.1 and $\sim$1.7\,MeV) is also affected by the
cosmogenic isotope  $^{11}$C ($\beta^+$ decay, Q\,=\,0.960\,MeV, $\tau$ = 29.4\,min) and by residual external background, mainly
as $\gamma$s from the decay of $^{208}$Tl (2.614\,MeV), $^{214}$Bi ($<$\,1.764\,MeV), and $^{40}$K (1.460\,MeV).

The $^{11}$C isotope is continuously produced in the liquid scintillator by muons through spallation on $^{12}$C. In order to limit its effect on the sensitivity
to $pep$ $\nu$s, we exploit the so-called Three-Fold Coincidence (TFC) method and $e^+/e^-$ pulse-shape discrimination~\cite{bib:Bxpep, bib:BxLong}.
 
\begin{figure}[h]
\begin{center}
 \includegraphics[width=0.5\textwidth]{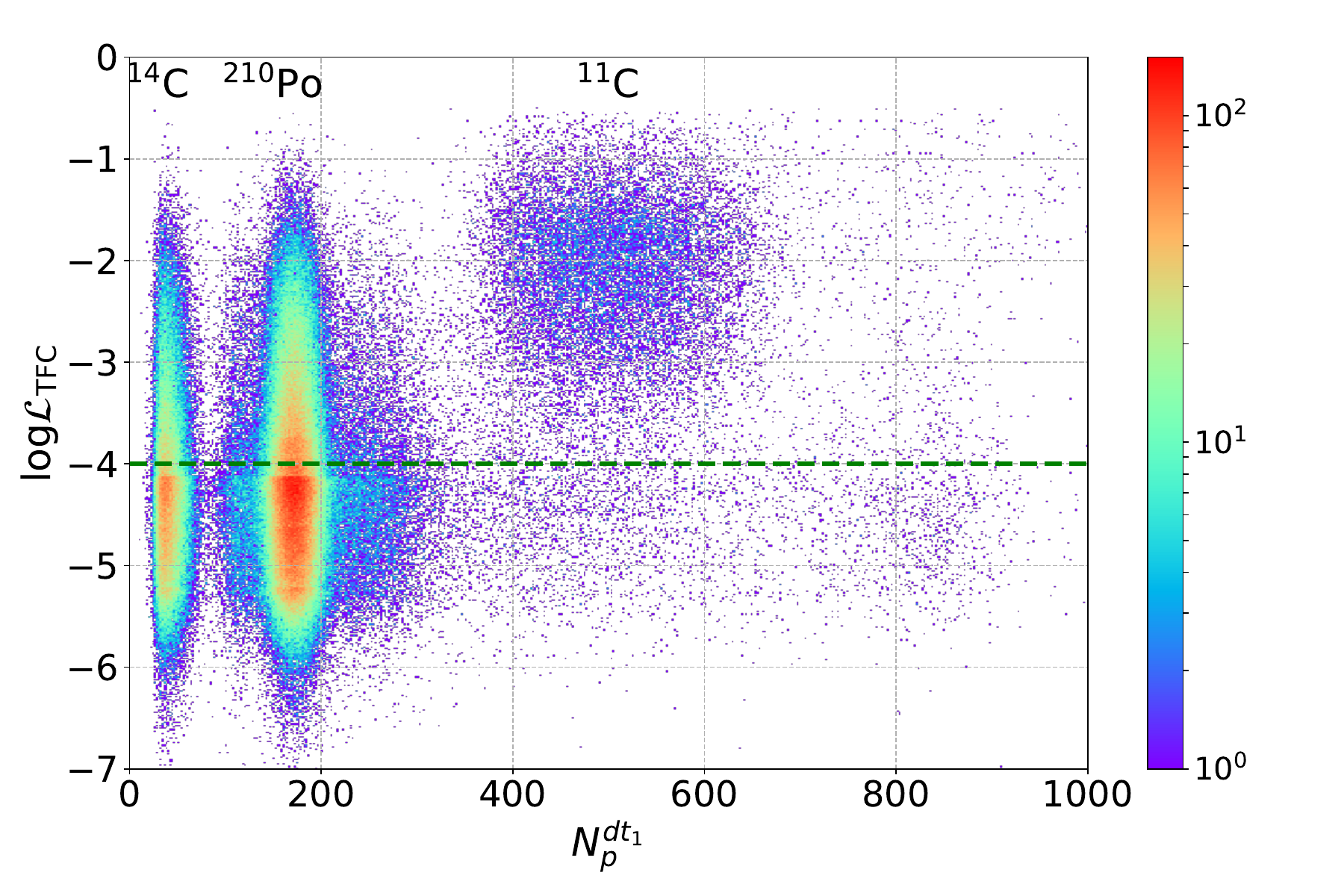}
 \caption{Distribution of log($\mathcal{L}_{\rm{TFC}}$) as a function of the $N_p^{dt_1}$  energy estimator. The plot is built using the entire set of
   data surviving the selection cuts described in Sec.~\ref{sec:det}. The regions dominated by the abundant internal background of $^{14}$C and $^{210}$Po are
   indicated by the corresponding labels. The green-dashed horizontal line represents the $\mathcal{L}_{\rm{TFC}}$-threshold, above/below which the events
   are assigned to the TFC-tagged/subtracted energy spectrum. It is clearly visible that the majority of the events of  the $^{11}$C energy decay spectrum lies above this threshold.}
\label{fig:C11Lik}
\end{center}
\end{figure}

The TFC takes advantage of the fact that  $^{11}$C is often produced together with one or even a burst of neutrons. 
The principle of the method  is thus to tag events correlated in space and time with a muon and a neutron.
We have improved the TFC technique already employed by us \cite{bib:Bxpep} by implementing a new algorithm, which evaluates  the likelihood $\mathcal{L}_{\rm{TFC}}$
that an event is a $^{11}$C candidate, considering relevant observables such as the distance in space and time from the parent muon, the distance from the neutron,
the neutron multiplicity, and muon $dE/dx$. 
Based on this probability, the data-set is divided in two samples: one depleted ({\it TFC-subtracted}), obtained removing the  $^{11}$C tagged events, and one enriched ({\it TFC-tagged}) in $^{11}$C. These two sets are
separately fitted in the multivariate scheme (see later).
The new TFC algorithm has (92\,$\pm$\,4)\% $^{11}$C-tagging efficiency, while preserving (64.28\,$\pm$\,0.01)\% of the total exposure in the TFC-subtracted spectrum.
Figure \ref{fig:C11Lik} shows the distribution of $\mathcal {\log({L}_{\rm{TFC}})}$ of the present data set as a function of the $N_p^{dt_1}$ energy estimator
and it demonstrates how $^{11}$C decays can be identified by cutting the events on the basis of the value of $\mathcal{L}_{\rm{TFC}}$.

\begin{figure}[h]
  \centering
  \includegraphics[width=0.48\textwidth]{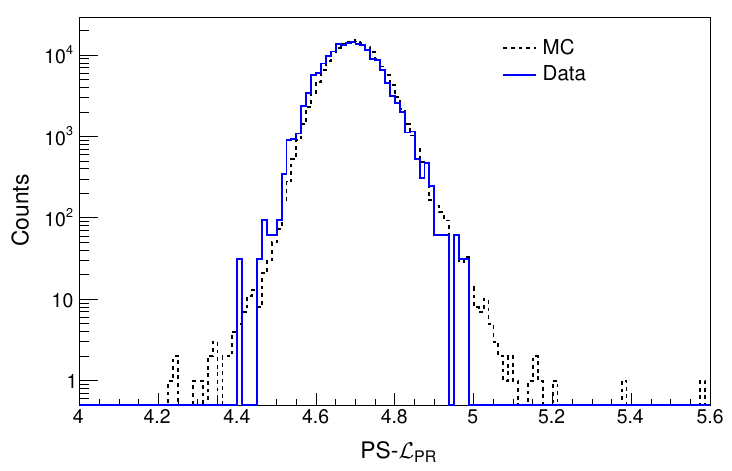}
  \caption{Comparison of the distributions of the PS-$\mathcal{L}_{\rm{PR}}$ parameter for $^{214}$Bi events extracted from data (blue, continuous line)
    and generated using MC (black, dashed line). The MC sample of $^{214}$Bi was generated using the same spatial distribution of the $^{214}$Bi events of the data.
    The simulation also takes into account the proper values of  the working channels $N^{\prime}(t)$.}
  \label{fig:userBi214}
\end{figure}

\begin{figure}[h]
  \centering
  \includegraphics[width=0.48\textwidth]{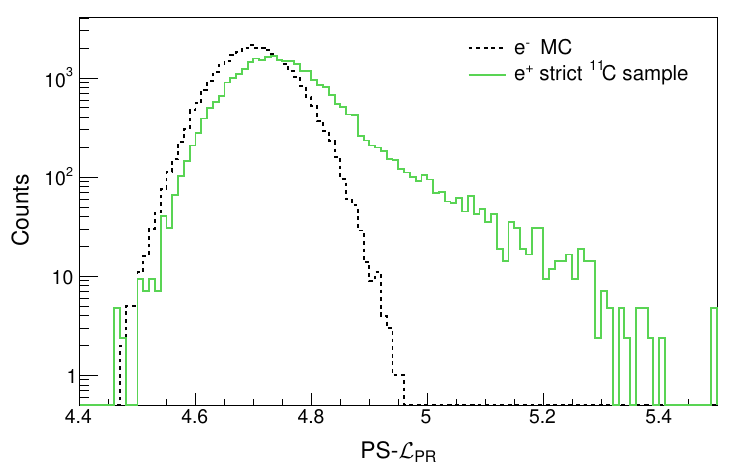}
  \caption{Comparison of the distributions of the PS-$\mathcal{L}_{\rm{PR}}$ variable for MC generated $e^-$ events (black, dashed line) and for $e^+$
    events selected from the data (green, continuous line). The latter events are a high-purity $^{11}$C sample, obtained with the optimized TFC method, using very strict
    cuts on the energy and on the time correlation with the neutron and muon tracks.}
  \label{fig:UserEleC11}
\end{figure}

\begin{figure}[h]
  \centering
  \includegraphics[width=0.5\textwidth]{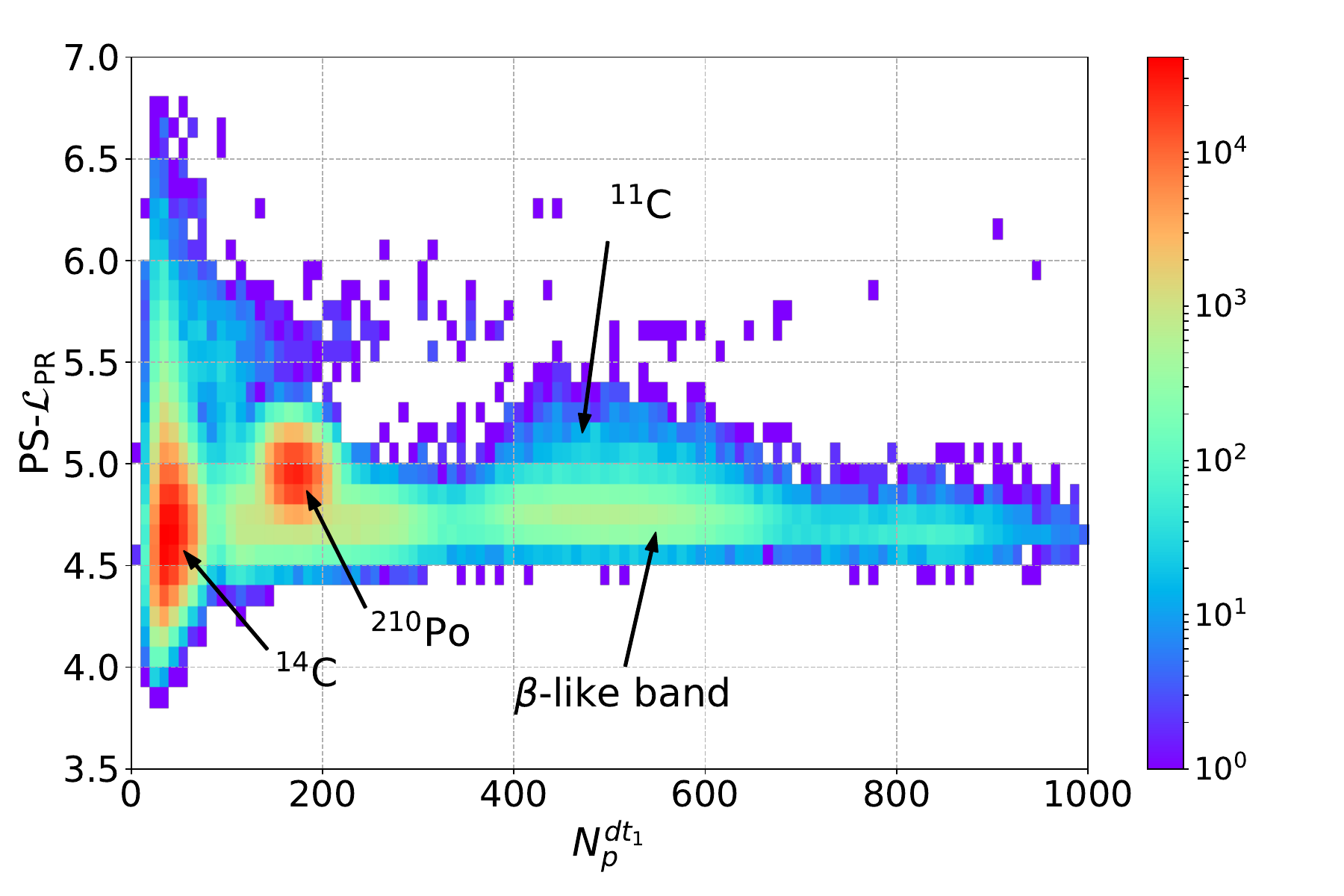}
  \caption{Distribution of PS-$\mathcal{L}_{\rm{PR}}$ pulse-shape discriminator as a function of $N_p^{dt_1}$ energy estimator. The plot is built using the
    entire set of data surviving the selection cuts described in Sec.~\ref{sec:det}.
The comparison with Fig. \ref{fig:UserEleC11} allows to identify the range of values belonging to the $\beta^--$like band indicated by the arrow.
  }
  \label{fig:userStCuts}
\end{figure}

\subsection{Pulse shape discrimination of  $\beta^+/\beta^-$events}
\label{subsec:PID}

The residual amount of $^{11}$C in the TFC-subtracted spectrum can be disentangled from the neutrino signal through variables with  $\beta^+/\beta^-$
pulse-shape discrimination capability~\cite{bib:Bxpep, bib:BxLong}. We build these variables considering that the  Probability Density Function (PDF) of the time
detection of the scintillation
light is  different for $\beta^+$ and $\beta^-$ events for two reasons: ${\it i)}$ for $\beta^+$ events, in 50\% of the cases, the $e^+$ annihilation is delayed by
ortho-positronium formation, which survives in the liquid scintillator with a mean time $\tau$\,$\sim$\,3\,ns~\cite{bib:DFranco}; ${\it ii)}$ the topology $e^+$ energy deposit
is not point-like, due to the two back-to-back 511\,keV annihilation $\gamma$s. 
These two features originate a pattern of the energy deposit of $\beta^+$ with a larger time and spatial spread than the corresponding one generated by $\beta^-$.
Based on this fact, a pulse--shape (PS) discrimination algorithm has been constructed using the neural network of a Boosted Decision Tree (BDT) and used for previous analysis
as detailed in~\cite{bib:BxLong}. In the present analysis, we have introduced a novel discrimination parameter, called PS-$\mathcal{L}_{\rm{PR}}$, defined as the maximum
value of the likelihood function $\mathcal{L}_{\rm{PR}}$ used in the position reconstruction (PR), divided by the value of the energy estimator. The latter normalisation removes
the $\mathcal{L}_{\rm{PR}}$ energy-dependence, since it is calculated as the summation over the collected hits~\cite{bib:BxLong}.
The PR-algorithm is based on the expected distribution of the arrival times of optical
photons on the PMTs. For all events, the algorithm uses the scintillation light emission
PDF  of point-like $\beta^-$ events.
For this reason the distribution of  the maximum likelihood value shows some discrimination capability for different types of particles, if they  originate
  photon time patterns distinct from that of  $\beta^-$.
  
The study of the performances of the PS-L$_{\rm{PR}}$ variable demands, from one side, the identification of  samples of true $\beta^-$ and $\beta^+$ events
and, from another side, it requires to properly account for the variable number of working channels that influences its value.

A pure, high-statistics $\beta^-$ sample can only be obtained from a limited time period of
the water-extraction phase of the scintillator purification campaign. During this time, a
temporary $^{222}$Rn contamination entered the detector. Using the space-and-time
correlation of the fast coinciding $^{214}$Bi($\beta^-$)-$^{214}$Po ($\alpha$) decays, we have tagged about $10^4$ $^{214}$Bi
events. The ability of the MC to reproduce the PS-L$_{\rm{PR}}$  parameter of these events and the comparison to data is shown in
 Fig.\,\ref{fig:userBi214}.
 The agreement between data and
 simulation demonstrates that the MC can accurately construct the PDF of this parameter for the entire set of data thus  accounting for the variable number of working channels.
 
 Our best  $\beta^+$ sample is  obtained from the TFC tagged events with hard cuts on the energy and on the time correlation with the neutron and muon tracks.
 These events are selected from the whole data set and thus they naturally follow the live-channels distribution.
The discrimination capability of  PS-L$_{\rm{PR}}$ is demonstrated comparing them with  a MC sample of pure
electrons with a flat energy distribution in the energy interval of the $^{11}$C events, while
also following the realistic live channel distribution over the whole data set. The  PS-L$_{\rm{PR}}$
 for these MC generated electrons was used as $\beta^-$ sample in a further analysis (analytical
multivariate fit) described in Sec.~\ref{sec:MVFit}). Figure~\ref{fig:UserEleC11}  shows the distribution of the PS-L$_{\rm{PR}}$
parameter for the MC generated electrons compared with that of $\beta^+$ events obtained
from $^{11}$C data.
The
difference between the two distributions at high values of PS-$\mathcal{L}_{\rm{PR}}$ is the key element allowing the discrimination between $\beta^-$ and $\beta^+$.
Note that we do not  need to build a position reconstruction algorithm based on  the time profile of the
scintillation light 
of $\beta^+$ events.
Figure~\ref{fig:userStCuts} shows the PS-$\mathcal{L}_{\rm{PR}}$ pulse-shape discriminator as a
function of $N_p^{dt_1}$ energy estimator for events selected with the cuts described in Sec. \ref{sec:det} and used in the present analysis.

It is interesting to note that the comparison between the  BDT and PS-$\mathcal{L}_{\rm{PR}}$ parameters, using the samples of true $\beta^-$ and $\beta^+$ events,
shows that they have similar discrimination power and they similarly help in reducing the systematic uncertainty of the $pep$ $\nu$s result.
However, the use of  PS-$\mathcal{L}_{\rm{PR}}$ offers some advantages like its simplicity, the fact that it can be
calculated without the training procedure necessary for BDT (that suffers the limited size of the available $\beta^-$ training sample), and finally, the possibility to
easily reproduce it through the MC.

\section{Multivariate fit}
\label{sec:MVFit}

The most powerful signatures for the detection of solar neutrinos in Borexino are the shapes of the energy spectra from electrons that underwent elastic scattering
interactions with neutrinos.
However, the  recognition of these shapes is somewhat obstacled by the contribution of various types of background events. In addition, the spectral details are also
masked through the finite energy resolution of the detector and eventually distorted by non-linear effects linking the energy deposit in the scintillator and the observed
energy estimator. 

Signal and background can be disentangled through an accurate fit. 
In order to enhance our sensitivity to the neutrino signal, we have adopted in the entire LER  the multivariate fit approach already exploited in~\cite{bib:Bxpep}.
We maximize a binned likelihood function containing the information from the TFC-tagged and TFC-subtracted energy
spectra. 
Additional information from the PS-$\mathcal{L}_{\rm{PR}}$ parameter and the radial distributions of the events in the optimized energy regions are included in the fit.
The radial information is important to accurately measure the background rates due to external $\gamma$s produced by the contamination of the PMTs 
and the supporting SSS.
The pulse shape parameter PS-$\mathcal{L}_{\rm{PR}}$ helps in the separation of the 
residual $^{11}$C($e^+$) background from the $e^-$--like components, and this
is relevant for the determination of $R_{pep}$ and $R_{\mathrm{CNO}}$.

Several ingredients are necessary to perform the  fit.  The first one is a background model, that is a list of possible radioactive contaminants  that we assume give a
contribution to the measured  signal.   The second one is the detector  response function, i.e.  a full model of the distributions of all the physical variables that we measure.
The knowledge of the detector response function allows the prediction of the  probability density functions of all the quantities entering  the fit procedure.

As done in previous Borexino analyses, we have adopted two complementary methods to build the detector response function:
an analytical approach
and a MC based procedure.
The only free parameters of the fit in the MC approach are the interaction rates of neutrino and background species, while in the analytical method (see later), in addition,
some of the parameters related to the response function  and to the energy scale  are also free and determined by the fit procedure.
These two methods share the same background model.

Fitting tools based on the use of Graphical Processing Units (GPU) have been developed and used with the analytical fit method. They decrease the computation time by
about 3 orders of magnitude compared with the 
standard CPU based algorithms previously used  \cite{bib:GPU}.

\subsection{Multivariate Likelihood Function}
\label{subsec:lkl}

The TFC-subtracted and TFC-tagged data-sets are fitted simultaneously by maximising a likelihood function $\mathcal{L}_{3D}(\vec{\theta}|\vec k)$ defined as 
\begin{equation}
  \mathcal{L}_{3D}(\vec{k} | \vec \theta) = \mathcal{L}^{TFC}_{sub}(\vec{k} | \vec \theta) \cdot
  \mathcal{L}^{TFC}_{tag}(\vec{k} | \vec \theta) .
  \label{eq:2Lik}
\end{equation}
The symbol $\vec{\theta}$ indicates the set of the arguments with respect to which the function is maximised and $\vec k$ generically indicates the set of the experimental
data used to evaluate the likelihood.
The two factors in Eq.~\ref{eq:2Lik} are the likelihood functions related to TFC-subtracted and TFC-tagged energy spectra, respectively.

$\mathcal{L}^{TFC}_{sub}(\vec{k} | \vec \theta)$  is the standard Poisson binned likelihood function:
\begin{equation}
  \mathcal{L}^{TFC}_{sub}(\vec{k} | \vec \theta) = \prod_{j,l,m=0}^{N_{E,R,P}}
  \frac{\lambda_{jlm}^{k_{jlm}}(\vec{\theta})}{k_{jlm}!}e^{-\lambda_{jlm}(\vec{\theta})},
  \label{eq:bxstatsLikelihood}
\end{equation}
where $\vec k$ in this case is the ensemble of    the data entries $k_{j,l,m}$ in the energy bin $j$, position bin $l$, and pulse shape parameter bin $m$;
$\lambda_{j,l,m}(\vec \theta)$ are the expected number
of entries in the same bins, and $N_{E,R,P}$ are  the total number of energy, radial, and pulse shape parameter bins.
 $\mathcal{L}^{TFC}_{tag}(\vec{k} | \vec \theta)$ is constructed in a similar way but it does not include  the pulse shape variable:
\begin{equation}
  \mathcal{L}^{TFC}_{tag}(\vec{k} | \vec \theta) = \prod_{j,l=0}^{N_{E,R}}
  \frac{\lambda_{j,l}^{k_{j,l}}(\vec{\theta})}{k_{j,l}!}e^{-\lambda_{j,l}(\vec{\theta})}.
  \label{eq:bxstatsLikelihood-tagged}
\end{equation}
and $\vec k$ represents in this case the set of data entries $k_{j,l}$  in the energy and radial bins $j,l$ integrated with respect to the pulse shape parameter.
The signal of $^{11}$C in the TFC-tagged spectrum is relatively strong compared to the other spectral components and the fit procedure extracts it very
  efficiently thanks to its spectral shape.
  This is the reason driving the choice of using the two dimensional (2D)  likelihood function of Eq. \ref{eq:bxstatsLikelihood-tagged}  for the TFC-tagged spectrum instead of a
  complete function of  Eq. \ref{eq:bxstatsLikelihood}
that, for the TFC-tagged spectrum, only increases the computation time without bringing additional information.

Both  the  TFC-subtracted and TFC-tagged spectra are fitted keeping the rates of the majority of the components in common, except $^{11}$C itself, $^6$He and $^{10}$C
(which have cosmogenic origin),
 and $^{210}$Po, that is not distributed homogeneoulsy through the detector volume.

Constraints on the values of the multivariate fit parameters are implemented (if not specified otherwise) as
multiplicative Gaussian terms in the likelihood function.

The likelihood function of Eq.~\ref{eq:bxstatsLikelihood} and Eq.~\ref{eq:bxstatsLikelihood-tagged} are exactly the ones which are
maximized using our most recent version of the MC-based fit
procedure (see Sec.~\ref{subsec:MCmethod}).
Precisely, we generate with the MC every signal and background component and we build and properly normalize 3D (or 2D) histograms of the simulated
  number of events as a function 
of the energy estimator,  PS-$\mathcal{L}_{\rm{PR}}$ parameter, and radius (or of the energy estimator and radius only).
The quantities $\lambda_{jlm}$ and $\lambda_{jl}$ of Eq.~\ref{eq:bxstatsLikelihood} and Eq.~\ref{eq:bxstatsLikelihood-tagged} represent
the sum of the bin content of the histograms,
each one weighted
by the  rate of the specific component ($\vec \theta$).

Earlier versions of the MC fit and the present analytical fit maximize an approximated
version of the likelihood  $\mathcal{L}_{3D}(\vec k| \vec \theta)$,  as already described in~\cite{bib:BxLong}. This function, called $\mathcal{L}(\vec{k} | \vec \theta)$,
is written as a product of four factors coming from the
TFC-subtracted and TFC-tagged energy spectra ( $\mathcal{L}^{TFC}_{E,sub}$ and $\mathcal{L}^{TFC}_{E,tag}$) 
and from the PS-$\mathcal{L}_{\rm{PR}}$  ($\mathcal{L}_P$) and radial  ($\mathcal{L}_R$) distributions
of events in the $^{11}$C-energy-range of the TFC-subtracted spectrum:

\vspace{-20pt}
\begin{center}
\begin{equation} 
  \mathcal{L}(\vec k | \vec \theta)=
  \mathcal{L}^{TFC}_{E,sub}(\vec k | \vec \theta)\cdot \mathcal{L}^{TFC}_{E,tag}(\vec k | \vec \theta)\cdot\mathcal{L}_P(\vec k| \vec \theta)\cdot\mathcal{L}_R(\vec k |\vec \theta).
\label{eq:LikeSimple}
\end{equation}
\end{center}

The first two terms,  $ \mathcal{L}^{TFC}_{E,sub}(\vec k| \vec \theta)$ and $ \mathcal{L}^{TFC}_{E,tag}(\vec k| \vec \theta)$, are  Poisson likelihoods
(like Eq.~\ref{eq:bxstatsLikelihood} and ~\ref{eq:bxstatsLikelihood-tagged})
with $\vec k$ being the data entries $k_j$ in the energy bin $j$ integrated with respect to the other variables.

The other two terms in  Eq.~\ref{eq:LikeSimple} have been built considering that in the framework of the analytical approach, there is no model able to produce precise multi--dimensional PDFs.
Thus we have  projected the events,
from the optimized energy intervals of the TFC-subtracted spectrum and integrated over energy ranges larger than the binning of the energy spectrum,
into 1D histograms of the pulse-shape and radial distributions.
$\mathcal{L}_P(\vec k | \vec \theta)$ and $ \mathcal{L}_R(\vec k| \vec \theta)$
of Eq.~\ref{eq:LikeSimple} are then built fitting these 1D distributions
using PDFs obtained either from the data (high purity $^{11}$C sample for $\beta^+$ pulse shape) or based on the MC simulation ($\beta^-$ pulse shape, radial distributions).
In the calculation of the corresponding likelihoods, we introduce a correlation between the number of counts in different histograms, as events that are in
the energy spectrum will also be entries in the projections.
To handle this issue, we normalise the functions to the total number of entries $N$ in the projected data histograms. Consequently, we define the
likelihood of the PS-$\mathcal{L}_{\rm{PR}}$ parameter as we did in~\cite{bib:BxLong} for the previously used PS-BDT parameter:
\begin{equation}
\mathcal{L}_P(\vec k | \vec \theta) = \prod^{N_P^{1D}}_{m=1} \frac{a\lambda_m^{k_m}(\vec{\theta})e^{-a\lambda_m(\vec{\theta})}}{k_m!}, 
\label{LBDT}
\end{equation}
 where the scaling parameter $a$ enforces the normalisation and is set such
\begin{equation}
 N = a \sum^{N_P^{1D}}_{m=1} \lambda_m(\vec{\theta}),
\label{eqn:adef}
\end{equation}
where $N$ is the total number of entries in the projected histogram and $a$ is a scaling factor.
Here, $k_m$ is the actual number of entries of bin $m$ of the 1D projection of the PS-$\mathcal{L}_{\rm{PR}}$ distribution in a fixed energy interval, $N_P^{1D}$
is the total number of bins of this histogram, and $\lambda_m(\vec{\theta})$ represents the expected content in bin $m$.
$\mathcal{L}_R(\vec k | \vec \theta)$ is defined in a way similar to $\mathcal{L}_P(\vec k | \vec \theta) $.

The results of the MC-based fit, which is either performed using
$\mathcal{L}_{3D}(\vec k | \vec \theta)$ or $\mathcal{L}(\vec k  | \vec \theta)$,
are consistent, confirming that no systematic uncertainty is introduced when using the approximated likelihood function.

\section{Detector response function}
\label{sec:response}

\subsection{The Monte Carlo method}
\label{subsec:MCmethod}

The MC code developed for Borexino~\cite{bib:BxMC} is a customised Geant4-based simulation package~\cite{bib:Geant4}, which can simulate all processes
following the interaction of a particle in the detector (energy loss including ionisation quenching in the scintillator; scintillation and Cherenkov light production; optical
photon propagation and interaction in the scintillator modelling absorption and re-emission, Rayleigh scattering,  interaction of the optical photons with the surface
of the materials; photon detection on the PMTs, and response of the electronics chain) including all known characteristics of the apparatus (geometry, properties
of the materials, variable number of the working channels over the duration of the experiment as in  the real data) and their evolution in time. The code thus produces a fully simulated detector  response function because it provides
a simulated
version of all the measured physical variables.
 
All the MC input parameters have been chosen or optimised using samples of data independent from the ones used in the present analysis (laboratory measurements
and Borexino calibrations with radioactive sources~\cite{bib:BxCalib}) and the simulation of the variables relevant for the present analysis has reached
sub-percent precision~\cite{bib:BxMC}.

Once the MC input parameters have been tuned, the PDFs of all the needed variables related to  each of the $\nu$ and background components are
built simulating events according to the specific energy spectrum. In order to properly  reproduce the spatial dependence of the energy response,
events are simulated in the detector following their 
expected spatial distribution: while the $\nu$ and most of background events are expected to be uniformly distributed in the detector, $^{210}$Po
decays are simulated according to their actual spatial and time distribution obtained from experimental data.
Note that data events due to the $\alpha$ decay of $^{210}$Po are efficiently identified by tagging $^{210}$Po with a pulse--shape discrimination method based on 
Multilayer Perceptron (MLP) algorithm~\cite{bib:BxSeas} (a particular class of neural network algorithms). Similarly, $\gamma$s from external
background are generated on the SSS and PMTs surfaces so that the radial distribution of the interactions inside the scintillator volume shows a clear decrease
from the outer region of the detector towards the center.

Events generated according to the theoretical signal and background energy spectra are then processed as real data. 
As already anticipated, for every species, 3D or 2D histograms are built for the energy estimators, the reconstructed radius, and the PS-$\mathcal{L}_{\rm{PR}}$ variable. When properly
binned and normalized, these histograms represent the PDFs  to be used in the fit and they provide the values $\lambda_{jlm}(\vec\theta)$ in Eq.~\ref{eq:bxstatsLikelihood}
and $\lambda_{jl}(\vec\theta)$ in Eq.~\ref{eq:bxstatsLikelihood-tagged}.
In the MC approach there are no free fit parameters other than the interaction rates of all species. The goodness of the fit
simultaneously demonstrates the accuracy of the MC simulation, as well as the stability of the detector response over the period of five years.
 
In the wide energy range covered by this analysis, there is a huge difference between the
number of measured counts per bin in the lower and in the higher energy regions. In the construction of the 3D PDFs, the need to simulate large numbers of events becomes
really important, since they are scattered over a larger number of bins. To mitigate the consequences due to low populated bins and to have a good approximation to a $\chi^2$,
we have replaced the energy estimator and the radius $R$ with some transformed variables. We choose to use $R^3$ instead of $R$, thus using bins of 5\,m$^3$
each and still achieving a very effective separation of the external background from the bulk components. Similarly, we
introduced a transformed variable $N_{h^\prime}$ based on the $N_h$ energy estimator: this change of variable is equivalent to adopting a variable binning
size that scales with energy proportionally to the width of the $N_h$ distribution obtained simulating mono-energetic electrons. This approach allows to reduce
the statistical fluctuations
without losing any physical information. As a by-product, this efficient binning significantly reduced the computing time needed to perform a single fit, speeding up  the
analysis of the MC pseudo--experiments   used to estimate the statistical and systematic uncertainties of the measurement described in Section~\ref{sec:sensitivity}.

 
The multivariate analysis was not applied on the whole energy range: the radial information was considered only for $N_h > 290$ to exclude from the analysis the spatial
distribution of $^{210}$Po, while the PS-$\mathcal{L}_{\rm{PR}}$ was used where $^{11}$C is present ($409 < N_h < 645$).
The shape of the probability density function of the PS-$\mathcal{L}_{\rm{PR}}$  variable for $\beta^+$ was obtained from an empirical parametrisation of the
distribution generated by the MC, with an additional small shift to compensate differences between the MC simulation outcome and a sample of strictly selected $^{11}$C events.

\begin{table*}
\small
\centering
\def\arraystretch{1.5}
\begin{tabular}{| C{1.9cm} |  C{1.7cm} | C{10cm} | C{3.0cm}|}
\hline
\textbf{Parameter}&\textbf{Fix./Free} & \textbf{Meaning/Approach to fixing} & \textbf{Value} \\
\hline
$Y^{pe}_0$& free & Photoelectron yield [p.e./MeV] for events in the detector center and with $N_{tot}$=2000 PMTs & 551 $\pm$ 1 \\
 \hline
$g_C$ & fixed & Fit $N_{p}^{dt_{1(2)}}$ vs  true $N_{pe}$ of MC  with Eq.~\ref{eq:np_npe} 
using MC mono-energetic electron samples at 4 energies, simulated along the whole data-set. &0.101   \\ 
\hline
$p_t$ & fixed & Fraction of a single photoelectron charge spectrum below the electronics threshold; fixed from the earlier  calibration measurements and calculations.
 & 0.12 \\
\hline
$f_{Ch}$ & fixed & Relative weight of the scintillation and Cherenkov light; fixed by performing many analytical fits on data with it as a free/fixed parameter.& 1.0 \\
\hline
$F_{Ch}(E)$ & fixed & 
                                  \begin{equation} 
                                  F_{Ch}(E)=  (C_0 +C_1 \cdot x + C_2 \cdot  x^2 +C_3 \cdot x^3 ) (1+C_4 \cdot E)
				  \end{equation}
				  $x=\ln{(1+E/E_0)}$;
				  
				  $E_0$ = 0.165 MeV & 
$C_0 = 1.415$; 
\vspace{2pt}
$C_1 = -3.397$;
\vspace{2pt}
$C_2 = 1.107$;
\vspace{2pt}
$C_3 = 0.072$;
\vspace{2pt}
$C_4 = 1.337$ 
\\
\hline
$Q(E)$ & fixed & 
Quenching term summarising the effects related to non-linearity of the scintillator response according to Birk's quenching model \cite{bib:BxLong}: 
\begin{equation}
Q(E,k_B)= \frac{1}{E}\int_0^E \frac{dE'}{1 +k_B dE'/dx},
\end{equation}
where $k_B$ is the Birk's constant, and Q(E) can be parametrised as:
\begin{equation}
Q(E,k_B)= \frac{A_1+A_2 \ln E + A_3 \ln E^2}{1+ A_4 \ln E + A_5  \ln E^2};
\end{equation}
fixed from the fit of $N_{pe}$  vs $E$ with MC simulation of $\gamma$ calibration data.
& 
\vspace{5pt}
$k_B[\rm{cm/MeV}] = 0.0109$; 
\vspace{5pt}
$A_1 = 0.972$; 
\vspace{5pt}
$A_2 = 0.201$;
\vspace{5pt}
$A_3 = 0.0105$;
\vspace{5pt}
$A_4 = 0.195$;
\vspace{5pt}
$A_5 = 0.014$
\\
\hline
 $v_1$ &   fixed &  Relative variance of the probability that a PMT triggers for events uniformly distributed in the detector volume, calculated using dedicated MC studies.
 It has some energy dependence and then  we are using a value averaged over the LER. &  0.16 \\
\hline
$v_T^0$ & free & Spatial non-uniformity of the number of triggered PMTs.  & 0.50 $\pm$ 0.37\\
\hline
$v_N$ &  free &  Scintillator intrinsic resolution parameter for $\beta$s (caused by $\delta$-electrons) that also effectively takes into account other contributions at low energies.
& 11.5 $\pm$ 1.0  \\
\hline
  $v^q_T$ & fixed & Non-uniformity of the light collection, calculated from MC events uniformly distributed in FV. &  7.0 \\
\hline
$v_T^{\alpha}$ &  free &  Spatial non-uniformity resolution,  
corresponding to the width of $^{210}$Po-$\alpha$ peak. & 4.73 $\pm$ 0.21\\
\hline
$\sigma_d$ & fixed & PMT dark noise contribution & 0.23  $N^{dt_1}_p$, 0.4  $N^{dt_2}_p$\\  
\hline
\end{tabular}
\caption{Summary of the parameters used in the analytical model of the energy response function. In case of the free parameters, the values given in the table
  are obtained with the multivariate analytical fit of the data discussed in this paper and reported in section \ref{sec:results}. Only statistical errors are  shown here.
  Energy is expressed in MeV;
 $v_T^0$, $v_T^{\alpha}$ ($v_T^q$) must be multiplied by  $10^{-6}$ ($10^{-4}$). }
\label{tab:bigTable}
\end{table*}

\subsection{The analytical model of the energy response function}
\label{subsec:analytical}

In the analytical approach, we introduce a PDF for the energy estimator under consideration and analytical expressions for its mean value and variance.
This PDF describes the detector's  energy response function  to  mono--energetic
events and, in brief,  it is mainly influenced by the number of scintillation and Cherenkov photons and effects due to the non--uniformity of the light collection.
As already anticipated,
we then transform the energy spectra of each species
 into the corresponding distributions of the energy estimators. 
 Effects like the ionisation quenching in the scintillator, the contribution of the Cherenkov light, the spatial dependence of the reconstructed energy and its resolution are
 accounted for through some parameters, part of which are fixed, while others are free to vary in the final fit. 
 
We describe here the present model  for  $N_p$ which 
 is derived from~\cite{bib:BxLong}, with several improvements to extend the energy range of the fit to the entire LER. The same model describes the variables $N_p^{dt_{1(2)}}$.
 All used energy estimators are obtained after normalising the corresponding measured values to a reference configuration of $N_{tot}=2000$ channels
 (defined in  Sec. \ref{sec:det}). 

  As energy response function for the entire LER, we use  the scaled Poisson function $f (N_p)$ (and similarly $f (N_p^{dt_{1(2)}})$)     already introduced for analyzing
  events in the lowest region of the energy spectrum and detailed in \cite{bib:TestCharge} and in \cite{bib:Oleg}
\begin{equation}
f (N_p) = \frac {m^{(N_p s)}}{(N_p s) !} \exp^{-m}
\label{eq:scaledP}
\end{equation}
The two  free parameters of this function,  $m$ and $s$, are fixed using the expressions for the mean value $\bar N_p(E)$ and variance $\sigma_p^2$ developed
in the context of  our model and described below:
\begin{equation}
m = \frac {\bar N_p^2(E)} {\sigma_p^2}
\end{equation}
and
\begin{equation}
s = \frac {\sigma_p^2} {\bar N_p(E)}
\end{equation}

In order to obtain $\bar N_p(E)$ we first consider that 
the mean number of photoelectrons $\bar N_{pe} (E)$ for each event of energy $E$ takes its main contribution from the scintillation photons with a sub-dominant
correction from the Cherenkov light, and it can be written as follows
\begin{equation}
\bar N_{pe}(E) = Y^{pe}_0 \cdot [Q(E) \cdot E  + f_{Ch} \cdot F_{Ch}(E)]
\end{equation}
where $Y^{pe}_0$ is the photoelectron yield expressed in photoelectrons/MeV for events in the detector center; 
the quenching term, $Q(E)$, accounts for  non-linearity of the scintillator response; $F_{Ch}(E)$, an  analytical parametrisation of Cherenkov light dependence on
energy valid for electrons,
provides the smooth transition between linear dependency at the energies above 1--2 MeV and zero contribution for electrons below the Cherenkov threshold
$E_0$ = 0.165\,MeV; $f_{Ch}$ is a parameter allowing to adjust the relative weight of the scintillation and Cherenkov light.
Table \ref{tab:bigTable} reports details of the analytical expressions.

Similarly to that described in \cite{bib:BxLong}, $\bar N_{p}(E)$, is linked to $\bar N_{pe}(E)$ through 
\begin{equation}
\bar N_{p} (E)= N_{tot}\left[1- e^{ -\mu} \left( 1+p_t  \mu\right) \right] \left( 1- g_C  \mu \right)
\label{eq:np_npe}
\end{equation}
where $\mu = \dfrac{\bar N_{pe}(E)}{N_{tot}}$, $g_C$ is a geometric correction factor, calculated for the given fiducial volume, and $p_t$
is the fraction of a single photoelectron signal below the electronics threshold.
These expressions extend the ones previously used in~\cite{bib:BxLong} with the introduction of the $f_{Ch}$ and $p_t$ parameters.

The second ingredient of the analytical model is the variance $\sigma^2_{p}$ of the $N_p$ energy estimator.  It is described  by  the following expression
which extends the model already described in ~\cite{bib:BxLong} in particular with the modification of the term  linear  with  $\bar N_p(E)$ and the addition of a quadratic one

%
%
%
\vspace{-5pt}
\begin{eqnarray}
\sigma^2_{p}  = f_{eq} \left[ 1- \left( 1+ v_1 \right)  p_1 \right] \bar N_{p}(E) + v_T^0 \bar N_{p}^3(E) +  \nonumber \\
+\   v^q_T {\left( \mu \frac{p_0}{p_1}\right)}^2 \bar N_{p}^2(E) + v_N  \bar N_{p}(E) +\sigma^2_d 
\label{eq:sigma_beta}
\end{eqnarray}
where $v_1$ is the relative variance of the PMT triggering probability for events uniformly distributed in the detector volume, 
$p_1 = 1 - e^{-\mu}$ is the probability of having a signal at any PMT, 
$p_0 = e^{- \mu}$ is the probability of absence of the signal,
$v_T^0$ accounts for the spatial non-uniformity of the number of triggered PMTs, 
 $v^q_T$ accounts for the non-uniformity of the light collection, 
$v_N$ is the intrinsic resolution parameter of the scintillator for $\beta$s 
that effectively includes other contributions at low energies, and the last term $\sigma_d$ describes the effects of the dark noise of the PMTs.
The channel equalization factor $f_{eq}$ is the ratio between $N_{tot}$ and the actual number of working PMTs and it changes during the data taking period.

In summary, the cubic term takes into account the variance of the number of the
triggered PMTs for the events with fixed collected charge in the IV. The
quadratic term takes into account the variance of the light collection
function over the detector, and is generally weaker compared to the
cubic term (and was neglected in previous analyses with more uniform
PMTs distribution).

Formula \ref{eq:sigma_beta} was derived analytically  and verified against the MC simulations.
For $\alpha$ particles
we are
using a simplified form with only the first and cubic term of relation \ref{eq:sigma_beta} since we need to model a   single energy point ($^{210}$Po).
It is thus not necessary to follow the energy dependence of the variance. The coefficient of the cubic term is called $v_T^{\alpha}$
and it corresponds to the width of the $^{210}$Po-$\alpha$ peak. 
As anticipated, we use the previous relations also for describing the mean value and variance of the estimators $N_p^{dt_{1(2)}}$.


Most of the above listed parameters are tuned using data independent from the ones used in the solar neutrino fit, calibrations or MC and are fixed in
the fit ($Q(E)$,  $f_{Ch}$, $p_t$, $g_C$, $v_1$, $v_T^{q}$), while
other parameters ($Y^{pe}$, $v_T^{\alpha }$, $v_T^{0}$,  $v_N$)  are left free to vary in the fit, together with the neutrino and background interaction rates.
The two parameters $p_t$, $g_C$ could be in principle free fit parameters, however they are fixed because  the fit results  have a low sensitivity to them.

In summary, the model has one free parameter describing the yield and three free parameters describing the energy resolution.
 Leaving the above listed parameters free gives the analytical fit the freedom to account 
 for  unexpected effects or unforeseen variations of the detector response in time. 
Table \ref{tab:bigTable} reports all the parameters, free or fixed,  appearing in the analytical fit with a short explanation about  how they are obtained. 
In case of parameters kept free in the fit, we report in the table the values obtained fitting the present data set as described in this paper. The corresponding values of
the $\nu$ interaction rates and 
background are reported in section \ref{sec:results}.

\subsection{Handling of the energy variables in the fit}
\label{sec:Oleg}
We perform the fit of the energy spectra with the experimental data binned as a function of the energy estimators instead of  trasforming that distributions into
the energy scale. Among the reasons driving this choice we  remark that 
the analytical approach does not assume  a priori knowledge of the precise energy transformation rules and
the energy scale is automatically adjusted while fitting the experimental data. The use of the transformed experimental spectra would significantly
slow down the fitting procedure, as 
the data reprocessing will be needed each time the energy scale parameters are changed in the fit. 
In addition,
  the presence of the contributions from $^{14}$C and $^{210}$Po with very high statistics  makes the fit sensitive to tiny  details of the energy response
  function (response to the
monoenergetic event with a fixed energy distributed uniformly in the detector's volume). The shape of 
the energy response for the detected number of p.e. (or the number of the triggered PMTs) in the sub-MeV energy region is 
defined mainly by the statistical factor, with small additional smearing due to the non-uniformity of the amount of the
collected light throughout the detector. The study performed using MC model showed that the shape of the charge response can be approximated
by the generalized gamma function, and the shape of the $N_p$ response can be approximated by the scaled Poisson function.
But the energy response function in the energy scale does not allow a simple description with an analytical function, and thus
complex calculations would be necessary if the transformed energy is used.
In the MC approach the transformation to the energy scale is in principle feasible,
because the energy scale and energy response in this approach are fixed from the calibrations, but it was not applied to keep internal consistency with the analytical approach.

Moreover, the amount of light emitted for a given energy deposit in the scintillator differs for the electrons, $\gamma$s and $\alpha$ particles and then the  
energy scale calibrated for electrons is not valid for $\alpha$s and $\gamma$s.
The
experimental spectrum contains contributions from all these types of particle and the event-by-event identification of the
type of interaction is not possible while  the different contributions are statistically identified  using the fit procedure.
The binning of the data in the physical energy scale (as shown in the figures reporting the fit results) is performed only after the fit is completed.

\section{Sensitivity studies}
\label{sec:sensitivity}

\begin{figure*}
\includegraphics[width = 1\textwidth]{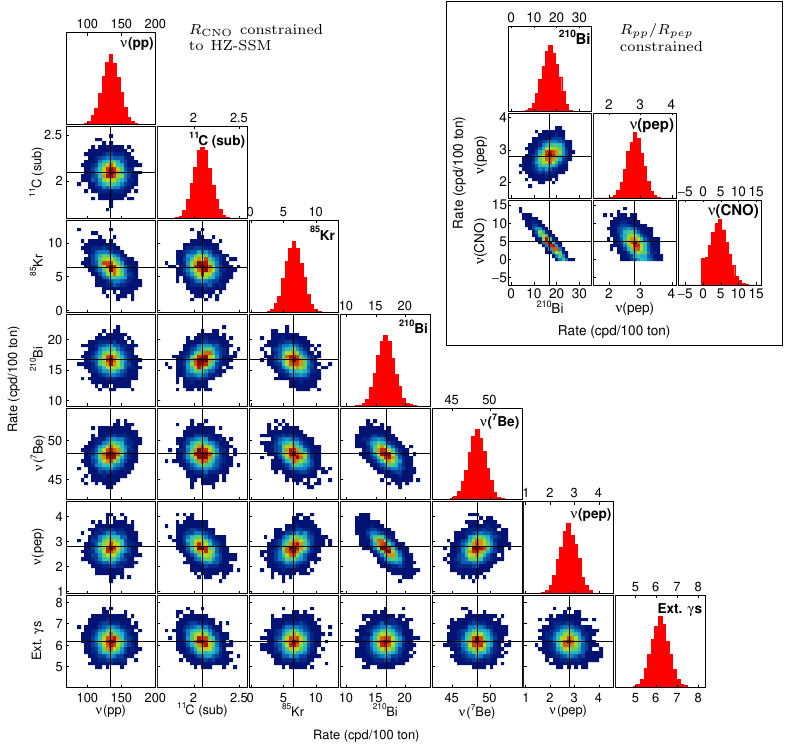}
\caption{The figure shows the distributions of the interaction rates (cpd/100\,ton) of solar $\nu$ and of the background species as they result from the MC fit of 
pseudo-experiments simulated with 
the same exposure as the experimental data discussed in this paper. The fit is performed in the entire LER region and, as in the real data analysis,
  penalty terms are added in the likelihood to constrain the values of the $^{14}$C and pileup rates within the  measured ones.
  It is interesting to note the correlation between the $pp$ and $^{85}$Kr rates, physically driven by the fact that a not negligible portion  of the $^{85}$Kr spectrum lies in the
  energy region around  about 200 keV where we are sensitive to the $pp$ $\nu$s signal. In the left plot, 6700 pseudo-experiments  have been generated
assuming the $R_{\rm CNO}$
according to HZ-SSM and fitted imposing a constraint on  $R_{\rm CNO}$ to the same value. The same MC PDFs have been used 
to simulate and fit data, so these plots show only uncertainties due to statistical fluctuations and the effects of the correlations among the various components.
The top right inset represents the results of the fit of  10000 pseudo-experiments fitted with the MC method while keeping the $R_{\rm CNO}$ free but constraining
the $R_{pp}/R_{pep}$ ratio to (47.7 $\pm$ 0.8) (HZ-SSM~\cite{bib:Carlos2017}, \cite{bib:Concha2017}). Constraining $R_{pp}$/$R_{pep}$ to the LZ-SSM prediction,
47.5 $\pm$ 0.8, gives consistent results. The study included all the background and neutrino species: here we only show those components that mostly influence
the sensitivity to CNO neutrinos.
}
\label{fig:CorrPlot}
\vspace{-0.2in}
\end{figure*}

Sensitivity studies have been performed by generating many pseudo-experiments with the MC
 and fitting this simulated data using the same response functions  adopted for fitting the real experimental data, using both analytical and MC procedures. 
 The simulated data of the pseudo-experiments are obtained from a random sampling of 
 PDFs produced with the full Borexino MC, including solar neutrino interaction rates as predicted by the HZ/LZ--SSM and with  the rates of the different background
 components   compatible with the final results presented in this work.
As an example, Fig.~\ref{fig:CorrPlot} shows the distribution of the results of the MC fit of 6700 pseudo-experiments each one with the same exposure as the real data.
In this particular example, by construction, the fit model perfectly matches the simulated data. The 1D distributions of the fit results, {\it i.e.} the rates $R$ of different solar
neutrino and background species, are Gaussian and do not show any significant biases with respect to the rates used as simulation inputs. The widths of these distributions
show the expected statistical precision of the measurement of the corresponding component. The shapes of the analogous 2D distributions visualize the correlations among
the different components. In particular, we underline that since the energy spectrum of the CNO neutrinos is quite similar to that of the $^{210}$Bi internal contamination
and the fit procedure cannot disentangle them, the sensitivity studies for all the $pp$-cycle neutrino and background components are performed by constraining the CNO rate.
These results are depicted in the left portion of Fig.~\ref{fig:CorrPlot} with  $R_{\rm CNO}$ generated and constrained assuming, as an example, the HZ--SSM. The same
constraint on $R_{\rm CNO}$ is used in fitting the real data, as it will be reported below.
Some additional significant correlations are present among some of the various species, as the figure is showing. This is one of the reasons why the best accuracy
in the determination of the interaction rates of solar neutrinos  is obtained by fitting the entire energy spectrum, as in the present analysis, thus best using
all the available information about details of the entire spectral shapes, instead of choosing partial energy regions.

The top right inset in Fig.~\ref{fig:CorrPlot}  demonstrates the sensitivity of the present data set to CNO neutrinos. In this case, no constraint on $R_{\rm CNO}$ is applied,
  but, to decrease the effect of the degeneracy of the spectral shapes, a constraint on the ratio between  $R_{pp}$ and $R_{pep}$, as expected from the SSM, is applied.
  It is interesting to note the strong anti-correlation between the $^{210}$Bi and CNO components which is originated by the above discussed similarities of their energy spectra.

\begin{figure}[t!]
\includegraphics[width = 0.5\textwidth]{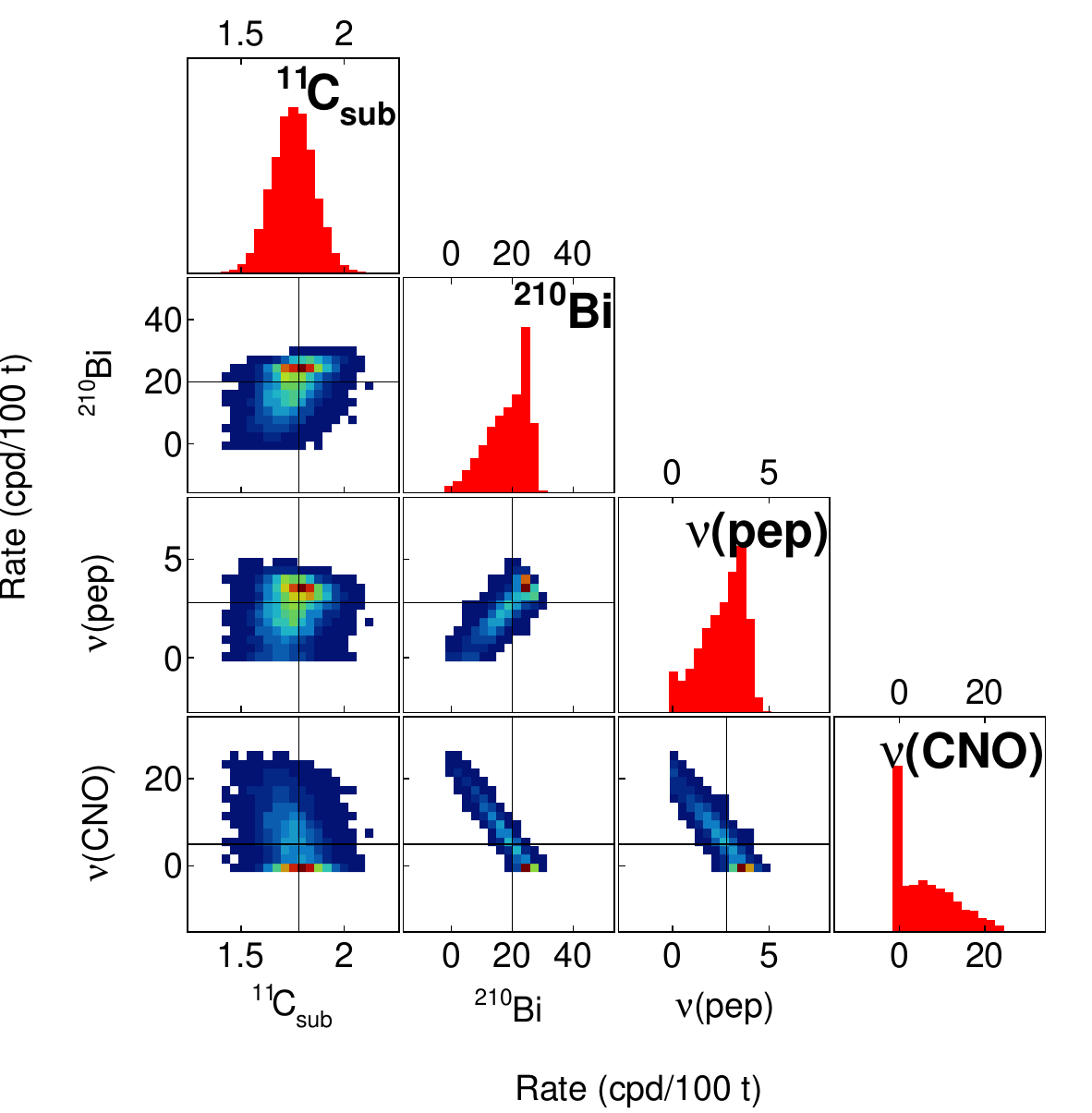}
\caption{
  This figure shows the results of the fit of MC simulated experiments obtained  in the same conditions of Fig. \ref{fig:CorrPlot} but, this time,
  removing all the constraints on $R_{\rm CNO}$ and $R_{pep}$.
  We only show here the correlation between $pep$, CNO, $^{11}$C and $^{210}$Bi, but the study included all the spectral components.
  The significant correlations among these species forbid to measure at the same time  $R_{\rm CNO}$ and $R_{pep}$ and to determine the    $^{210}$Bi decay rate.
  As described in the text, we have constrained the CNO rate to get the $pep$ one and set a constraint on the  ratio $R_{pp}$/$R_{pep}$ to obtain a limit on the CNO flux. }
  \label{fig:CorrPlot2}
\vspace{-0.2in}
\end{figure}

  Finally,  Fig.~\ref{fig:CorrPlot2} is obtained removing all the constraints on the CNO and $pep$ components and clearly shows that the strong correlations
    (and anti-correlations) among
   $R_{\rm CNO}$, $R_{pep}$, and the $^{210}$Bi decay rate significantly limit the possibility to determine all the three species at the same time.

   Similar MC studies have been performed to quantify the systematic uncertainty associated to the fit models, by generating MC data with a response function
   modified with respect to the one used in the fit (see next section). Finally, pseudo-experiments MC data have been used to obtain the distribution
   of the likelihood functions and thus evaluate the p-values of our results.
 
\begin{figure*}
\begin{center}
\begin{minipage}[c]{\textwidth}
\centering{\includegraphics[width = 0.48\textwidth]{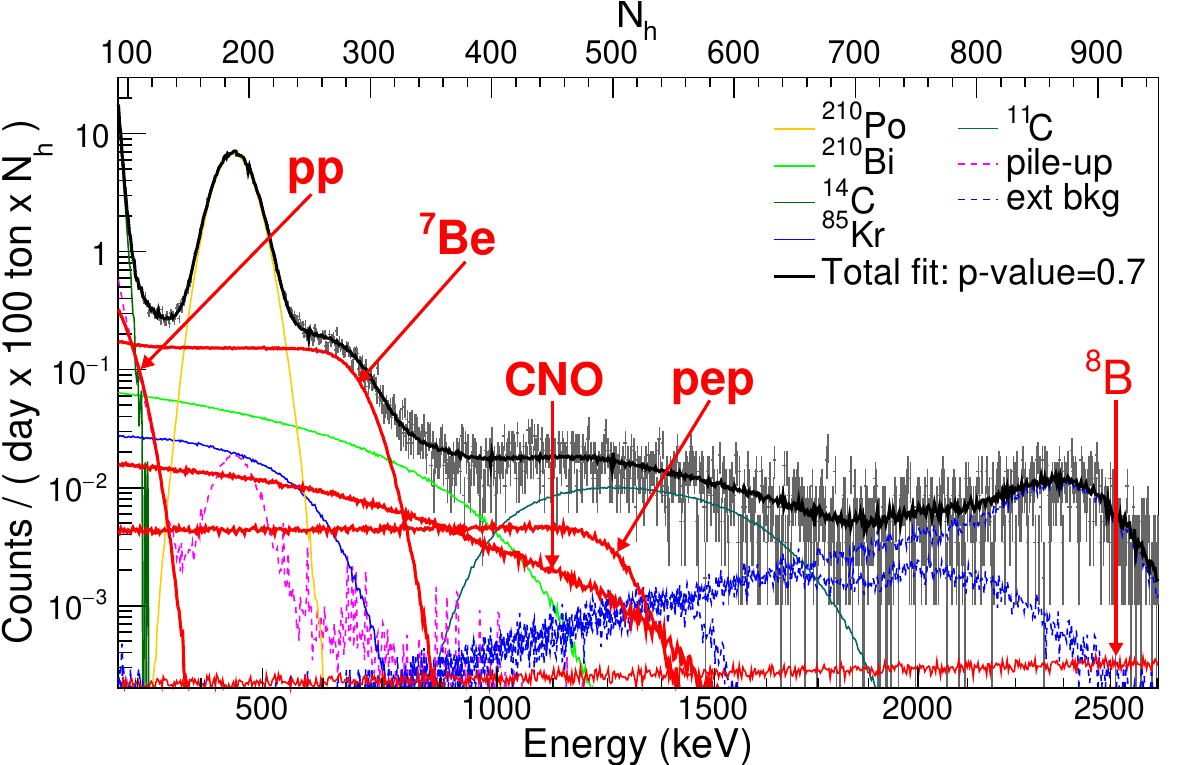}}
\hfill
\centering{\includegraphics[width = 0.48\textwidth]{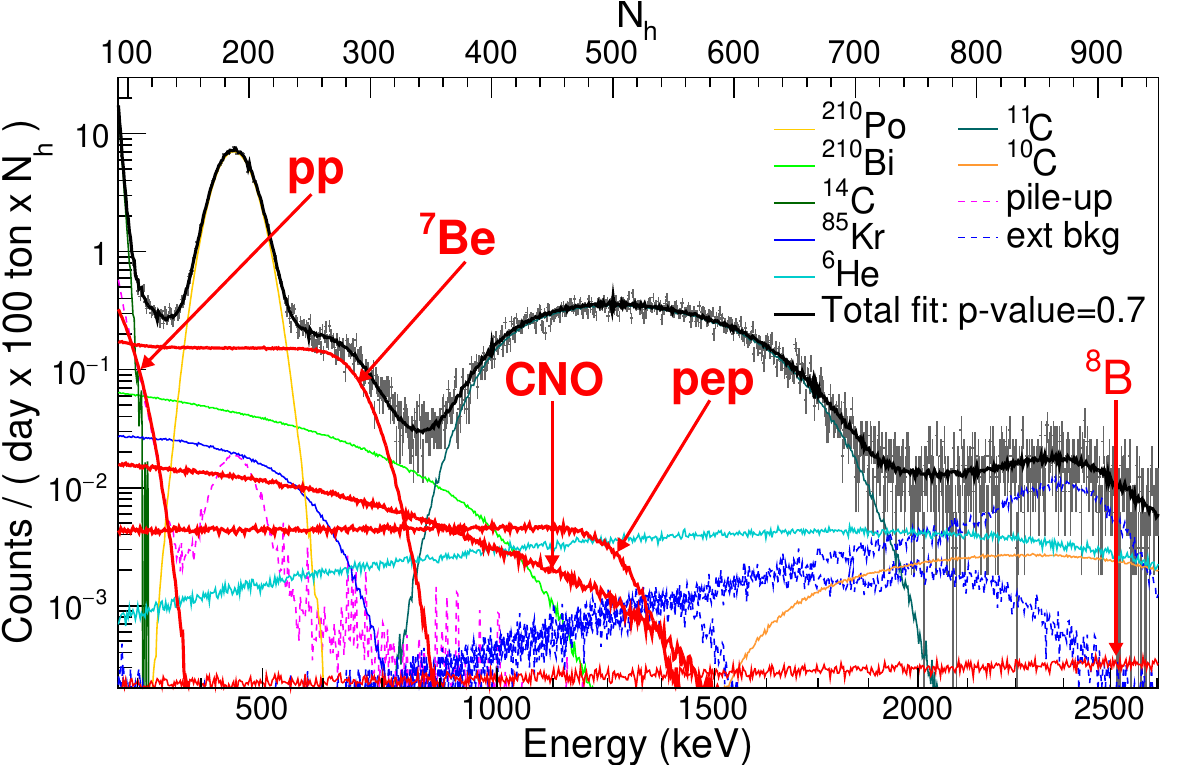}}
\vspace{-0.3em}\\
\centering{\includegraphics[width = 0.48\textwidth]{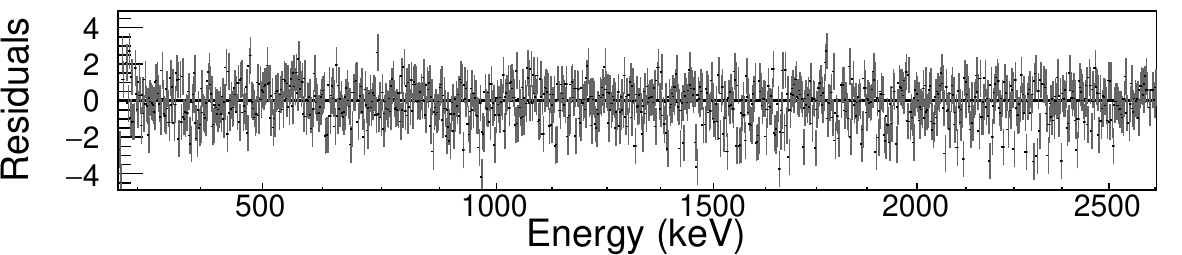}}
\hfill
\centering{\includegraphics[width = 0.48\textwidth]{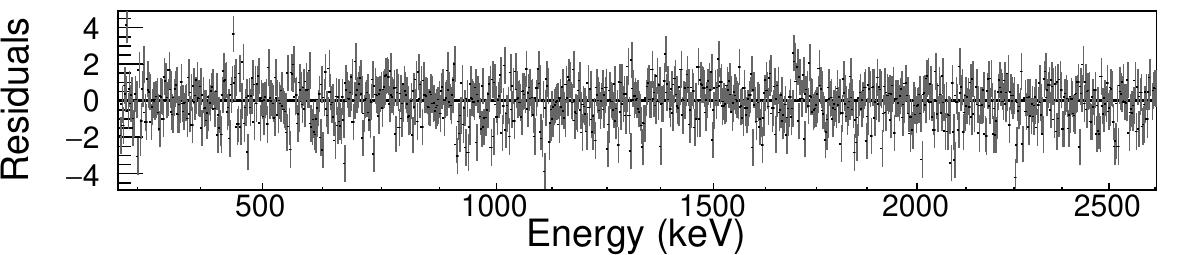}}
\caption{Multivariate fit results (an example obtained with the MC method) for the TFC-subtracted (left) and the TFC-tagged (right) energy spectra, with residuals.
  The sum of the individual components from the fit (black lines) is superimposed on the data (grey points). The analysis has been performed using $N_h$ as energy
  estimator and the transformation to keV-energy scale was perfomed only for the plotting purposes. The residuals are calculated in
  every bin as the difference between
the data counts and the fit result, divided by the square root of the data counts.}
\label{fig:SpectrumFit}
\end{minipage}
\end{center}
\vspace{-0.1in}
\end{figure*}

\begin{figure}
\begin{center}
\centering{\includegraphics[width = 0.45\textwidth]{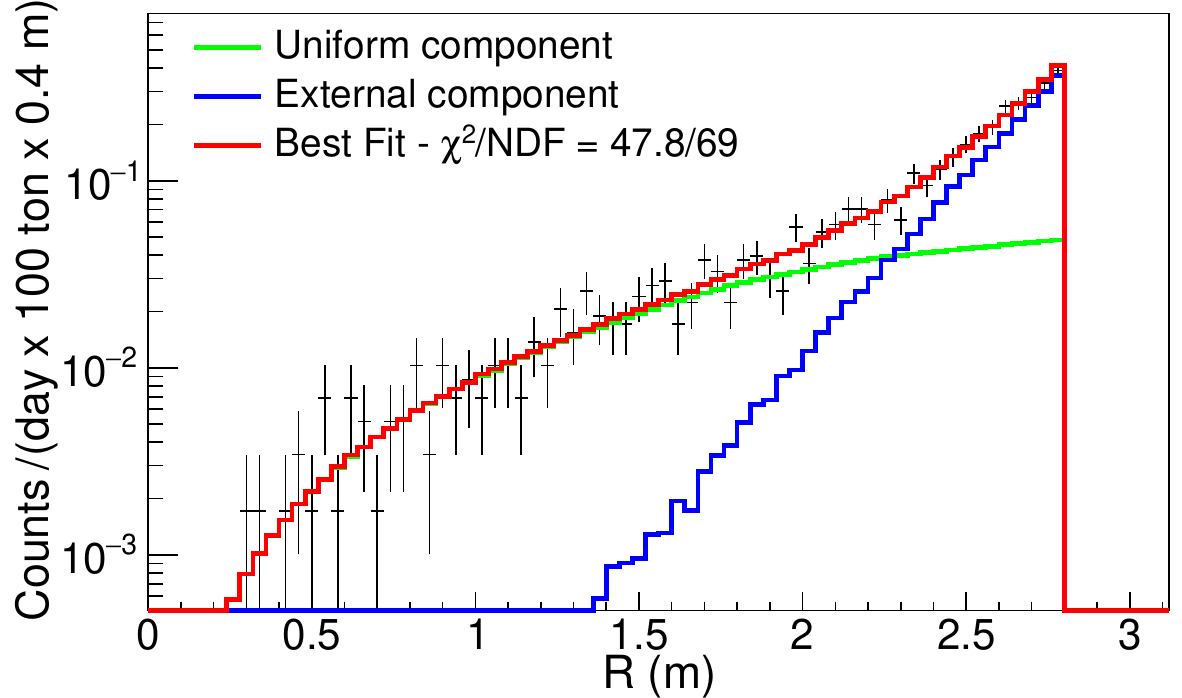}}
\centering{\includegraphics[width = 0.45\textwidth]{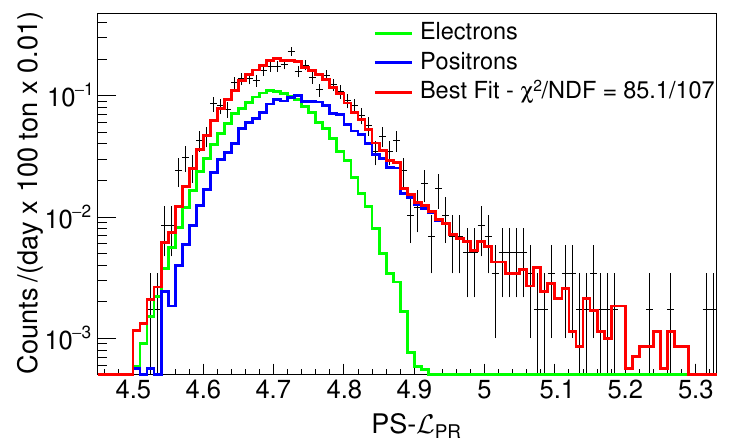}}
\caption{An example of the multivariate fit showing the radial (top) and the PS-$\mathcal{L}_{\rm{PR}}$ (bottom) distributions of the events (black crosses)
  from the TFC-subtracted spectrum, in the 
 energy intervals  $N_h > 290$ and $409 < N_h < 645$, respectively.}
\label{fig:user_radial}
\end{center}
\vspace{-0.1in}
\end{figure}

\begin{figure}
\centering\includegraphics[width = 0.48\textwidth]{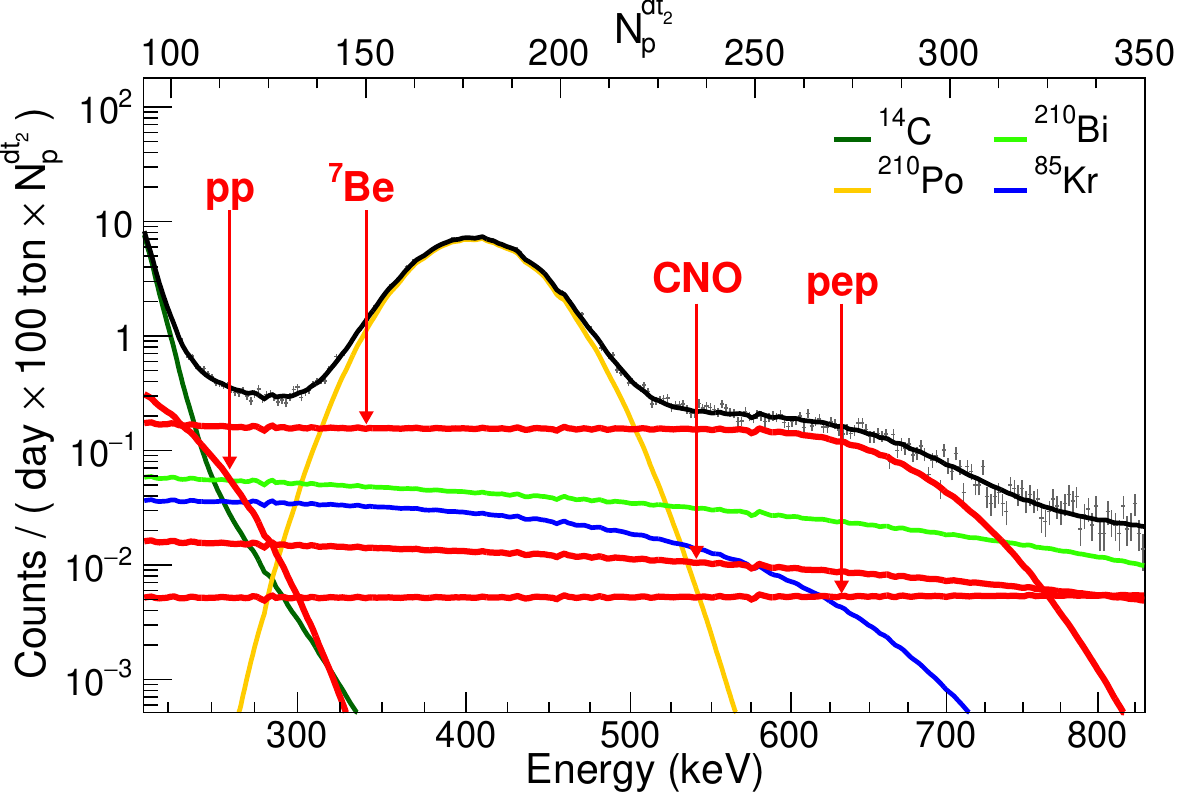}
\vspace{-1.3em}\\
\centering\includegraphics[width = 0.48\textwidth]{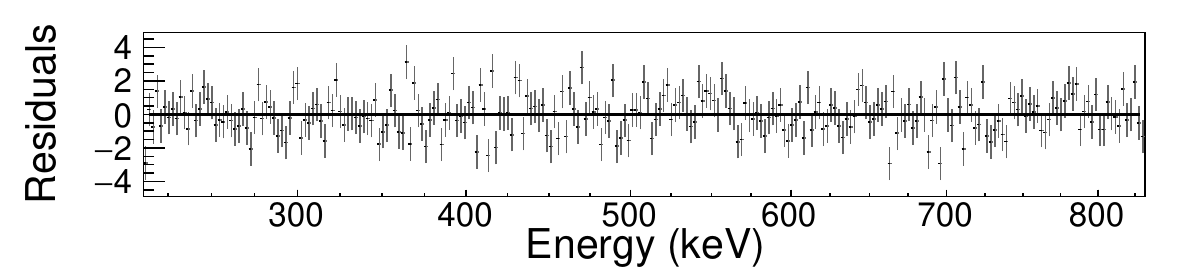}
\caption{Results of the fit for TFC-subtracted energy spectrum zoomed in to the lowest energy region (an example obtained with the analytical method) and residuals.
The residual are calculated as in Fig. \ref{fig:SpectrumFit}.} 
\label{fig:LowEnergy}
\vspace{-0.2in}
\end{figure}

\begin{table*}
\as{1.3}
\begin{center}
\begin{tabular}{l|c|c|c|c|c|c}
\hline \hline
\multirow{3}{*}{Solar $\nu$} & \multicolumn{2}{c|}{Borexino experimental results}  & \multicolumn{2}{c|}{B16(GS98)-HZ} & \multicolumn{2}{c}{B16(AGSS09)-LZ}\\
\cline{2-7}
    & Rate  & Flux  & Rate & Flux  & Rate  & Flux \\[-1ex]
    &  \footnotesize{[cpd/100\,ton]} &   \footnotesize{$\rm{[cm^{-2}s^{-1}]}$} &   \footnotesize{[cpd/100\,ton]} &   \footnotesize{$\rm{[cm^{-2}s^{-1}]}$}  &   \footnotesize{[cpd/100\,ton]}  & \footnotesize{$\rm{[cm^{-2}s^{-1}]}$} \\
\hline
$pp$  &
$134\pm10\;^{+6}_{-10}$   & $(6.1\pm0.5\;^{+0.3}_{-0.5})\times 10^{10}$ & 
$131.1 \pm 1.4$ & $5.98\,(1\pm0.006)\times10^{10}$ &
$132.2 \pm 1.4$ & $6.03\,(1\pm0.005)\times10^{10}$ \\
\hline
$\rm{^7Be}$  &
$48.3\pm1.1\;^{+0.4}_{-0.7}$  & $(4.99\pm0.11\;^{+0.06}_{-0.08})\times 10^{9}$ & 
$47.9 \pm 2.8$ & $4.93\,(1\pm0.06)\times10^{9}$ &
$43.7 \pm 2.5$ & $4.50\,(1\pm0.06)\times10^{9}$ \\
\hline
$pep$ (HZ)  &
$2.43 \pm 0.36\;^{+0.15}_{-0.22}$  &$(1.27\pm0.19\;^{+0.08}_{-0.12})\times 10^{8}$&
$2.74 \pm 0.04$ & $1.44\,(1\pm0.009)\times10^{8}$ &
$2.78 \pm 0.04$ & $1.46\,(1\pm0.009)\times10^{8}$ \\
\hline
$pep$ (LZ)  &
$2.65 \pm 0.36\;^{+0.15}_{-0.24}$  &$(1.39\pm0.19\;^{+0.08}_{-0.13})\times 10^{8}$&
$2.74 \pm 0.04$ & $1.44\,(1\pm0.009)\times10^{8}$ &
$2.78 \pm 0.04$ & $1.46\,(1\pm0.009)\times10^{8}$ \\

\hline
$\rm{CNO}$  &
$< 8.1\;(95\%\, \rm{C.L.})$  & $< 7.9\times10^8\;(95\%\, \rm{C.L.})$& 
$4.92 \pm 0.55$ & $4.88\,(1\pm0.11)\times10^{8}$ &
$3.52 \pm 0.37$ & $3.51\,(1\pm0.10)\times10^{8}$ \\
\hline
\hline
\end{tabular}
\end{center}

\vspace{-0.2in}
\caption{
  Borexino Phase-II results on $pp$, $^7$Be (862 +384 keV), $pep$, and CNO solar $\nu$s: interaction rates and fluxes inferred assuming the  MSW-LMA oscillation
  parameters~\cite{bib:Concha2017}.
The first error is the statistical derived by profiling the likelihood under Wilks' approximation. The interval extracted is consistent with the expectation from the MC sensitivity
study. The second error is the systematic uncertainty. Different contributions to the systematic error are detailed in Table\,\ref{tab:syst}.
 The result on $pep$ $\nu$s depends on whether we assume HZ-SSM or LZ-SSM metallicity for CNO $\nu$s. 
 The remaining columns show the theoretical interaction rates and fluxes  predicted by the Standard Solar Model under the high and low metallicity
 assumptions~\cite{bib:Carlos2017}.}
\label{tab:NuResults}
\end{table*}

\begin{table}
\as{1.1}
\begin{center}
\begin{tabular}{lcc}
\hline
\hline
Background	& ~~~ & ~~Rate \\[-0.5ex]
			& & \footnotesize{~~[cpd/100\, ton]}	    	\\
\hline
$^{14}$C~\footnotesize{[Bq/100~t]}	&	& $~40.0 \pm 2.0$  \\
$^{85}$Kr		&	& $~~6.8 \pm 1.8 $	  \\
$^{210}$Bi	&	& $~17.5 \pm 1.9 $	 	 \\
$^{11}$C  	 	&	& $~26.8 \pm 0.2$		 \\
$^{210}$Po 	&	& $260.0\pm 3.0$  \\
Ext. $^{40}$K 	&	& $~~1.0 \pm 0.6 $	\\
Ext. $^{214}$Bi	&	& $~~1.9 \pm 0.3 $	\\
Ext. $^{208}$Tl	&	& $~~3.3 \pm 0.1 $		\\
\hline
\hline
\end{tabular}
\end{center}
\vspace{-0.2in}
\caption{
  Best estimates for the total rates of the background species included in the fit with statistical and systematic uncertainties added in quadrature.
  These numbers are obtained by averaging the results of the fits  with the HZ and LZ hypothesis. }
\label{tab:bkg}

\end{table}

\begin{table}
\as{1.2}
\begin{center}
\begin{tabular}{l|cc|cc|cc}
\hline
\hline
                                                              
&\multicolumn{2}{c|}{$pp$}& \multicolumn{2}{c|}{$^7$Be} & \multicolumn{2}{c}{$pep$}  \\ 

\cline{2-7}							    
Source of uncertainty & $-\%$	& $+\%$	& $-\%$	& $+\%$	& $-\%$	& $+\%$ \\
\hline

Fit method (analytical/MC)                   &-1.2    &1.2              &-0.2         & 0.2     &-4.0  &4.0\\
Choice of energy estimator                        &-2.5   &2.5                 &-0.1        & 0.1     &-2.4 &2.4 \\
Pile-up modeling		                         &-2.5   &0.5                 &0        & 0     &0 & 0 \\
Fit range and binning                                &-3.0   &3.0                &-0.1       & 0.1      &1.0 &1.0 \\
Fit models (see text)                                 &-4.5  & 0.5                     &-1.0     &  0.2    &-6.8  &2.8\\
Inclusion of  $^{85}$Kr constraint 	                       &-2.2   &2.2               &0       & 0.4      &-3.2  &0\\

Live Time                                       &-0.05     &0.05           &-0.05    & 0.05     &-0.05  &0.05\\
Scintillator density                         &-0.05    &0.05               &-0.05      & 0.05     &-0.05  &0.05\\
Fiducial volume                                        &-1.1    &0.6              &-1.1       &  0.6       &-1.1 &0.6 \\
\hline                 
Total systematics ($\%$)		&-7.1	&4.7                   &-1.5      &0.8     &-9.0 &5.6\\
\hline
\hline
\end{tabular}
\end{center}
\vspace{-0.2in}
\caption{
Relevant sources of systematic uncertainty  and their contributions to the measured neutrino interaction rates.
More details are in the text.}
\label{tab:syst}

\end{table}

\begin{figure}
\includegraphics[width = 0.48\textwidth]{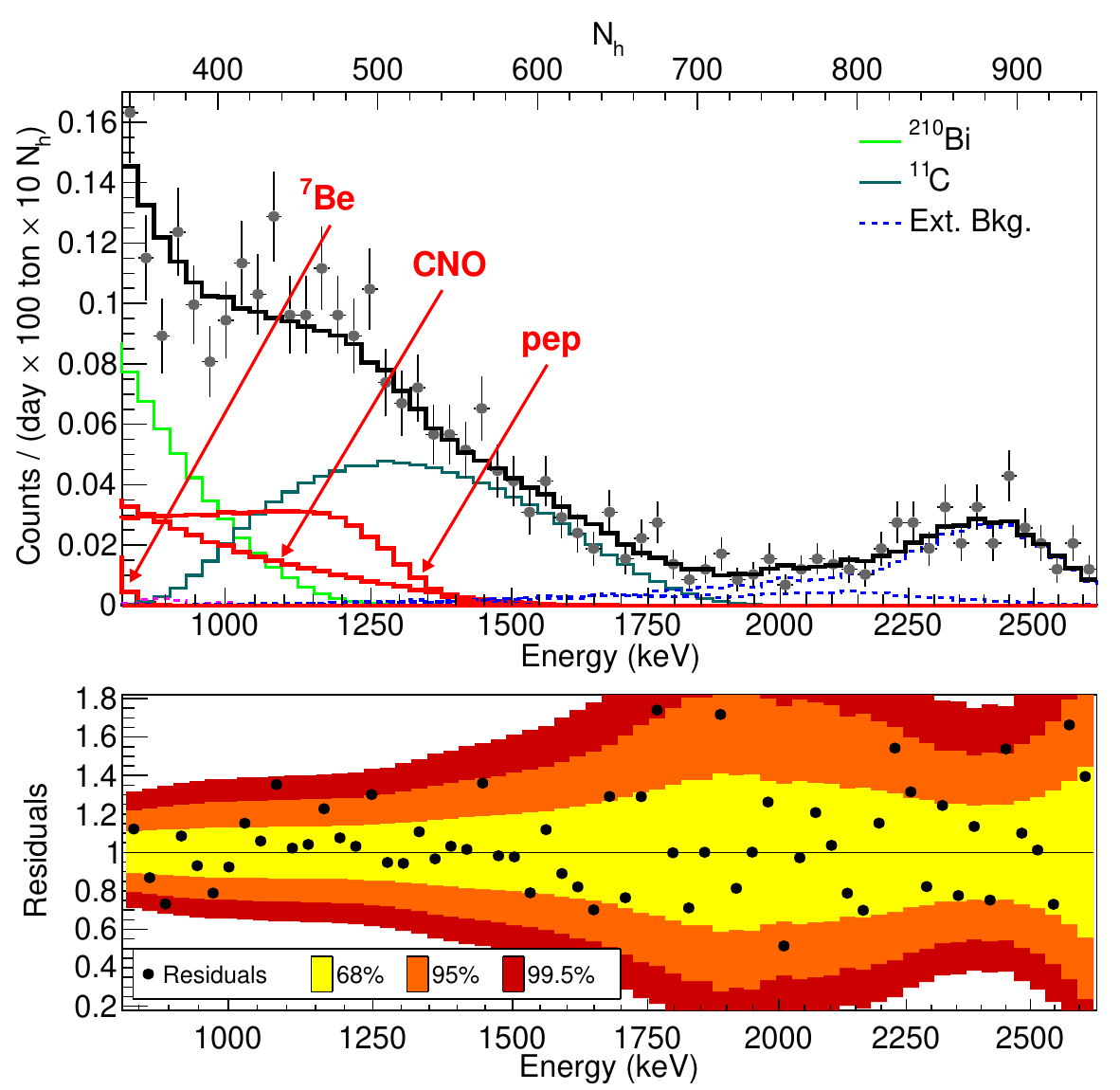}
\caption{TFC-subtracted energy spectrum zoomed between 800\,keV and 2700\,keV after applying  stringent selection cuts on the radial
  distribution (R\,$\,<\,$\,2.4\,m) and on the pulse-shape variable distribution (PS-$\mathcal{L}_{\rm{PR}}<\,$4.8) to better see features due to $pep$ $\nu$s interactions.
  The residuals (bottom plot) are the ratio between the data and the fit model.}
\label{fig:pep}
\vspace{-0.2in}
\end{figure}

\section{Results}
\label{sec:results}

The interaction rates $R_{pp}$, $R_{\rm{Be}}$, $R_{pep}$ are obtained from the fit together with the decay rates of 
 $^{85}$Kr, $^{210}$Po, $^{210}$Bi, $^{11}$C  internal backgrounds, and the external backgrounds rates ($^{208}$Tl, $^{214}$Bi, and $^{40}$K $\gamma$ rays). 

In the MC approach, the MC-based pile-up spectrum~\cite{bib:BxMC} is included in the fit with a constraint of (137.5\,$\pm$\,2.8~cpd/100\,ton)
on the $^{14}$C--$^{14}$C contribution based on an independent measurement of the  $^{14}$C rate~\cite{bib:Bxpp}.
In the analytical approach, pile-up is taken into account with the convolution of each spectral component with the solicited-trigger spectrum~\cite{bib:Bxpp}.
Alternatively, the analytical fit uses a synthetic pile-up spectrum~\cite{bib:Bxpp} built directly from data.
The differences between these methods are quoted in the systematic error (see Table~\,\ref{tab:syst}).

In order to break the degeneracy between the $^{210}$Bi and the CNO $\nu$ spectral shapes, we constrain the CNO $\nu$ interaction rate to the HZ-SSM predictions,
including MSW-LMA oscillations (4.92 $\pm$ 0.56 cpd/100\,ton)~\cite{bib:Carlos2017, bib:Concha2017} as anticipated in Sec.~\ref{sec:sensitivity}.
 The analysis is repeated constraining the CNO $\nu$ rate to the LZ-SSM predictions (3.52 $\pm$ 0.37 cpd/100\,ton) and in case of difference, the two results are quoted separately. 
 The contribution of $^8$B $\nu$s is small and its rate was constrained to the
value obtained from the HER analysis \cite{bib:BxB8Phase2}.

The interaction rates of solar neutrinos and the decay rates of background species, obtained by averaging the results of the analytical and MC approaches,
are summarised in Tables~\ref{tab:NuResults} and~\ref{tab:bkg}, respectively.

An example of the multivariate fit (with the MC approach) is shown in Fig.\,\ref{fig:SpectrumFit} (TFC-subtracted and TFC-tagged energy spectra) and in Fig.\,\ref{fig:user_radial}
(radial distribution and PS-$\mathcal{L}_{\rm{PR}}$ pulse-shape distribution).
The details of the fit at low energies (between $\sim$\,230 and 830\,keV) can be appreciated in Fig.~\ref{fig:LowEnergy}. In this example, obtained with the analytical
fit procedure, the pile-up is not present  as a separate fit component, since it is taken into account with the convolution method mentioned above.

To recognise the $pep$ $\nu$ contribution to the measured electron-recoil spectrum, the TFC-subtracted spectrum, zoomed into the highest energy region
(between 800 and 2700\,keV), is shown after applying stringent selection cuts on the radial distribution (R\,$<$\,2.4\,m) and on the pulse-shape variable
distribution (PS-$\mathcal{L}_{\rm{PR}}$ $<$4.8) (see Fig.\,\ref{fig:pep}): 
the CNO and pep neutrino interactions are clearly
visible between 1250 and 1500 keV, and the spectrum is consistent with the
Compton-like  shoulder expected from the $pep$ line.

An extensive study of the systematic errors has been performed and the results are summarised in Table~\ref{tab:syst}.

Differences between the results of the analytical and the MC fits are quoted as systematic errors. 
Further systematic  uncertainties associated with the fitting procedure  were studied by performing the fit in many different configurations
by generating simulated data using a family of response functions whose parameters has been varied within calibration accuracy with respect to the nominal response
function and by varying the energy estimator, the number and width of the bins, as well as the fit range). 

Systematic uncertainties related to the fit models  were evaluated using the method described in Sec.~\ref{sec:sensitivity}. Ensembles of pseudo-experiments
were generated from a family of  PDFs based on the full MC simulations and fitted using both the MC and analytical methods. PDFs
including deformations due to possible inaccuracies in the modeling of the detector response (energy scale, uniformity of the energy response,
shape of PS-$\mathcal{L}_{\rm{PR}}$) and uncertainties in the theoretical energy spectra ($^{210}$Bi) were considered. The magnitude of the deformation
was chosen to be within the range allowed by the available calibration data.

In an additional systematic study, the fit was repeated taking into account the upper limit on the $^{85}$Kr decay rate following the procedure
described in~\cite{bib:BxLong}, which exploits the $^{85}$Kr -- $^{85m}$Rb delayed coincidences ($^{85}$Kr rate $<$\,7.5\,cpd/100\,ton at 95\% C.L.).

The last three lines of Table\,\ref{tab:syst} list the uncertainties associated with the determination of the exposure. The one about  the fiducial volume is one of
the dominant.
  Its value is the  same as quoted in~\cite{bib:BxBe7} and it  is estimated using calibration sources of known positions.

  Fully consistent results are obtained when adopting a larger fiducial volume (R$\,<\,$3.02\,m, $\lvert z \rvert$\,$<$\,1.67\,m). This FV contains more external background
  (critical for the $pep$ $\nu$s) which is, however, properly disentangled by the multivariate fit thanks to its energy shape and radial distribution. The previously published
  Borexino results regarding $pp$ $\nu$s~\cite{bib:Bxpp} and $^7$Be $\nu$s~\cite{bib:BxBe7} were obtained in this enlarged fiducial volume.

  Finally, the analytical fit performed on a restricted energy range (not sensitive to $pp$ neutrinos) using the $N_{pe}$ energy estimator gives consistent results
  (within 2\,$\sigma$) for $R_{\rm{Be}}$ and $R_{pep}$. 

The $^7$Be solar $\nu$ flux listed in Table~\ref{tab:NuResults} is the sum of the two mono-energetic lines at 384 and 862\,keV. It corresponds to a rate for
  the 862\,keV line of 46.3\,$\pm$\,1.1$^{+0.4}_{-0.7}$\,cpd/100\,ton, fully compatible with the Borexino Phase-I measurement~\cite{bib:BxBe7}.
  The $^7$Be solar $\nu$ flux is determined with a total uncertainty of 2.7\,$\%$, which represents a factor of 1.8 improvement with respect to our previous
  result~\cite{bib:BxBe7} and is two times smaller than the theoretical uncertainty. 

The present value of  $R_{pp}$ is consistent with our previous result  and the uncertainty is  reduced by about 7.9\,\%.

The correlation between the CNO and $pep$ $\nu$ is broken by constraining the $R_{\rm{CNO}}$ in the fit. 
The values of $R_{\rm{Be}}$ and $R_{pp}$  are not affected by the hypothesis on CNO $\nu$s within our sensitivity. However,  $R_{pep}$ 
depends on it, being 0.22\,cpd/100\,ton higher if the LZ hypothesis is assumed (see Table\,\ref{tab:NuResults}). 

\begin{figure}
\begin{centering}
\includegraphics[width = 0.48\textwidth]{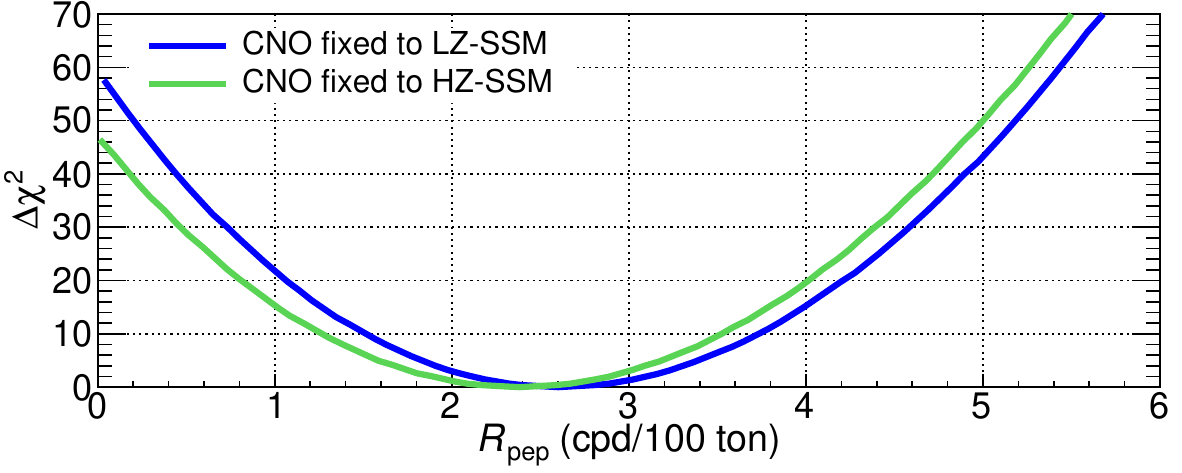}
\includegraphics[width = 0.48\textwidth]{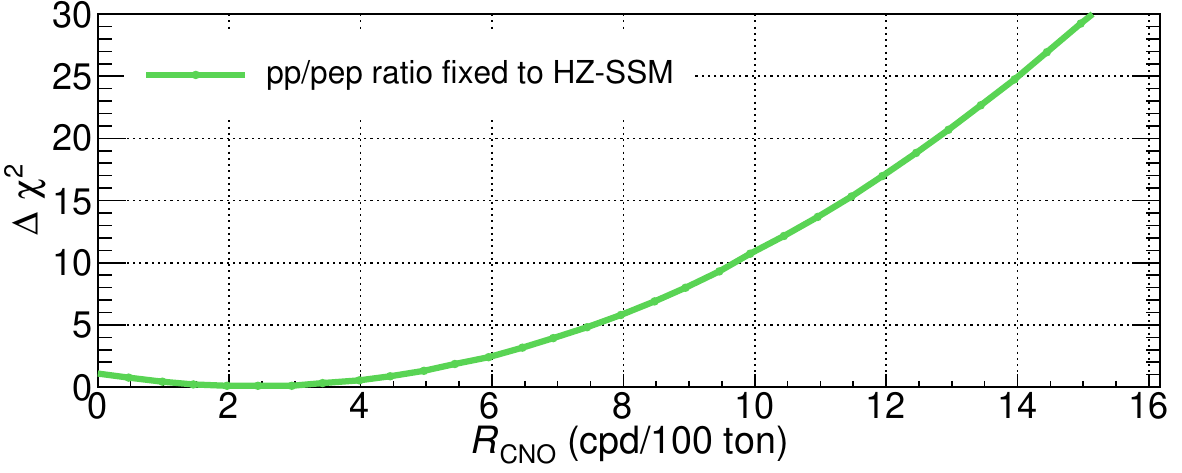}
\caption{$\Delta \chi^2$ profile for the $pep$ (top plot) and CNO (bottom plot) $\nu$ interaction rates.}
\label{fig:pep_CNO}
\end{centering}
\end{figure}

The $\Delta \chi^2$ profile obtained by marginalising the $pep$ rate
is shown in Fig.\,\ref{fig:pep_CNO} (left)   for both the HZ and LZ assumptions on CNO $\nu$ rate. Both curves are symmetric and allow us to establish,
for the first time, that the absence of the $pep$ reaction in the Sun is rejected at more than 5\,$\sigma$. 

As anticipated, the similarity between the $e^-$ recoil spectrum induced by the CNO neutrinos and the $^{210}$Bi spectrum makes it impossible to disentangle
the two contributions with the spectral fit  without an external constraint on the $^{210}$Bi rate. For this reason, we can only provide an upper limit on the CNO neutrinos
interaction rate $R_{\rm{CNO}}$. In order to do so, we need further to break the correlation between the CNO and $pep$ contributions. In Phase-I,
this was achieved by fixing the $pep$ $\nu$ rate to the theoretical value~\cite{bib:Bxpep}. In the current analysis,
where $pp$ $\nu$s are included in the extended energy range of the fit, we place an indirect constraint on $pep$ $\nu$s by exploiting the theoretically
well known $pp$ and $pep$ flux ratio. The interaction rate ratio $R_{pp}$/$R_{pep}$, is constrained to \,(47.7 $\pm$ 0.8) (HZ) \cite{bib:Carlos2017},
\cite{bib:Concha2017}.  Constraining $R_{pp}$/$R_{pep}$ to the LZ hypothesis value 47.5 $\pm$ 0.8 gives identical results. 

We carried out a  sensitivity study by performing the analysis on thousands  of data-sets simulated with the MC sensitivity  tool: this study shows that under the
current experimental conditions the total expected uncertainty (statistical plus systematical) is 3.4\,cpd/100\,ton. With this error,
we expect the median 95\% C.L. upper limit for $R_{\rm{CNO}}$ to be $\sim$\,9\,cpd/100\,ton and  10\,cpd/100\,ton, for low and high metallicity, respectively. 
On data, we obtain the upper limit on $R_{\rm{CNO}}$ = 8.1\,cpd/100\,ton (95 $\%$ C.L.) (see Table\,\ref{tab:NuResults}), which is slightly stronger
than the median limit expected from the MC based sensitivity study. The $\Delta \chi^2$  profile for the CNO rate is shown in Fig.~\ref{fig:pep_CNO} (bottom).
This result, using a weaker hypothesis on $pep$ $\nu$, confirms the current best limit on the flux of CNO $\nu$s   previously obtained
with Borexino Phase-I data~\cite{bib:Bxpep}.

\section{Conclusions}
In summary, we have reported the details of the analyis and the results of  the first simultaneous measurement of the $pp$, $^7$Be, and $pep$ components
of the solar neutrino spectrum providing a comprehensive investigation of the main $pp$ chain in the Sun  \cite{bib:BxNaturePhase2}.
These  results are in agreement with and improve the precision of our previous measurements.
 In particular, $R_{Be}$  is measured with an unprecedented precision of 2.7\%.
  The absence of $pep$ neutrinos is rejected for the first time at more than 5$\,\sigma$.
 These data, together with our measurement about $^8$B $\nu$ flux in the HER~\cite{bib:BxB8Phase2},
provide a unique measurement of the interaction rates and thus of the  fluxes of the  different components of the solar neutrinos   from the $pp$ chain with a
single detector and a unified analysis approach.



The upper limit  on $R_{\rm{CNO}}$ 
has the same significance as that of Borexino Phase-I  and currently is providing the tightest bound on this component.

Several  analysis methods and details here reported and discussed have a general interest which is going beyond the understanding of the Borexino results:
as example the $^{11}$C suppression,
  the multivariate fit, the analytical model of the energy response, the full MC description of the detector  and the fitting procedures can be easily adapted
  to large volume liquid scintillator based detectors similar to Borexino
 \cite{bib:Juno}, \cite{bib:SNOPlus}.

\section{Acknowledgement}
The Borexino program is made possible by funding from INFN (Italy), NSF (USA), BMBF, DFG, HGF, and MPG (Germany),
RFBR (Grants 16-02-01026 A, 19-02-0097A, 16-29-13014 ofim, 17-02-00305 A)(Russia), RSF (grant 17-12-01009) (Russia),
 and NCN (Grant No. UMO 2013/10/E/ST2/00180-Grant No. UMO 2017/26/M/ST2/00915) (Poland). 
We acknowledge  the computing services of Bologna INFN-CNAF data centre and LNGS Computing and Network Service (Italy), of Jülich Supercomputing Centre at FZJ
(Germany), and of ACK Cyfronet AGH Cracow (Poland). We acknowledge also the generous
hospitality and support of the Laboratori Nazionali del Gran Sasso (Italy).

\Urlmuskip=0mu plus 1mu\relax

\nocite{*}
\bibliography{NuSolPaperAfterNature}
\bibliographystyle{apsrev4-1}
\end{document}